\documentclass[aps,pra,twocolumn,superscriptaddress,floatfix,longbibliography,letter]{revtex4-1}
\usepackage[bookmarks=true,colorlinks,linkcolor=OrangeRed,urlcolor=NavyBlue,citecolor=RoyalBlue]{hyperref}
\usepackage{epsfig,amsmath,amssymb,color,dsfont,upgreek,physics}
\usepackage{subfiles}
\usepackage{mathrsfs}
\usepackage{mathtools}
\usepackage{bbold}
\usepackage[makeroom]{cancel}
\usepackage[caption=false]{subfig}
\usepackage{mathrsfs}
\usepackage{graphicx}
\usepackage{subfig}
\usepackage[countmax]{subfloat}
\usepackage[english]{babel}
\usepackage[dvipsnames]{xcolor}

\usepackage{braket}

\definecolor{mypurple}{rgb}{0.49,0.18,0.56}
\definecolor{mygold}{rgb}{0.93,0.69,0.13}
\definecolor{mygreen}{rgb}{0,0.5,0}
\definecolor{myblue}{rgb}{0,0,0.75}
\definecolor{mymagenta}{cmyk}{0,1,0,0.12}
\definecolor{mygray}{rgb}{0.5,0.5,0.5}

\usepackage{umoline}

\begin{document}

\title{Disorder-free localization with Stark gauge protection}
\author{Haifeng Lang}
\affiliation{INO-CNR BEC Center and Department of Physics, University of Trento, Via Sommarive 14, I-38123 Trento, Italy}
\affiliation{Theoretical Chemistry, Institute of Physical Chemistry, Heidelberg University, Im Neuenheimer Feld 229, 69120 Heidelberg, Germany}
\author{Philipp Hauke}
\affiliation{INO-CNR BEC Center and Department of Physics, University of Trento, Via Sommarive 14, I-38123 Trento, Italy}
\author{Johannes Knolle}
\affiliation{Department of Physics, Technische Universit\"at M\"unchen, James-Franck-Straße 1, D-85748 Garching, Germany}
\affiliation{Munich Center for Quantum Science and Technology (MCQST), Schellingstra\ss e 4, D-80799 M\"unchen, Germany}
\affiliation{Blackett Laboratory, Imperial College London, London SW7 2AZ, United Kingdom}
\author{Fabian Grusdt}
\affiliation{Department of Physics and Arnold Sommerfeld Center for Theoretical Physics (ASC), Ludwig-Maximilians-Universit\"at M\"unchen, Theresienstra\ss e 37, D-80333 M\"unchen, Germany}
\affiliation{Munich Center for Quantum Science and Technology (MCQST), Schellingstra\ss e 4, D-80799 M\"unchen, Germany}
\author{Jad C.~Halimeh}
\email{jad.halimeh@physik.lmu.de}
\affiliation{Department of Physics and Arnold Sommerfeld Center for Theoretical Physics (ASC), Ludwig-Maximilians-Universit\"at M\"unchen, Theresienstra\ss e 37, D-80333 M\"unchen, Germany}
\affiliation{Munich Center for Quantum Science and Technology (MCQST), Schellingstra\ss e 4, D-80799 M\"unchen, Germany}
\begin{abstract}
Disorder-free localization in translation-invariant gauge theories presents a counterintuitive yet powerful framework of ergodicity breaking in quantum many-body physics. The fragility of this phenomenon in the presence of gauge-breaking errors has recently been addressed, but no scheme has been able to reliably stabilize disorder-free localization through all accessible evolution times while preserving the disorder-free property. Here, we introduce the concept of \textit{Stark gauge protection}, which entails a linear sum in gauge-symmetry local (pseudo)generators weighted by a Stark potential. Using exact diagonalization and Krylov-based methods, we show how this scheme can stabilize or even enhance disorder-free localization against gauge-breaking errors in $\mathrm{U}(1)$ and $\mathbb{Z}_2$ gauge theories up to all accessible evolution times, without inducing \textit{bona fide} Stark many-body localization. We show through a Magnus expansion that the dynamics under Stark gauge protection is described by an effective Hamiltonian where gauge-breaking terms are suppressed locally by the protection strength and additionally by the matter site index, which we argue is the main reason behind stabilizing the localization up to all accessible times. Our scheme is readily feasible in modern ultracold-atom experiments and Rydberg-atom setups with optical tweezers.
\end{abstract}
\date{\today}
\maketitle
\tableofcontents

\section{Introduction}
Far-from-equilibrium quantum many-body dynamics is a prime application of today's quantum simulators \cite{Hauke2012,Georgescu_review,Altman_review}. With the great level of control and precision achieved in these setups \cite{Greiner2002,Bloch2008,Bakr2009}, fundamental phenomena in many-body dynamics are now being explored that had been restricted to the theoretical realm just a few years ago \cite{Schreiber2015,Smith2016,Choi2016,Jurcevic2017,Flaeschner2018,Kaplan2020}.

Among the most fascinating of these is many-body localization (MBL), which has originally been shown to result from the interplay of interactions and spatial disorder in a system (disorder-MBL) \cite{Basko2006,Alet_review,Abanin_review}. This paradigm of ergodicity breaking has been the focus of a large body of work, with impressive experimental observations \cite{Kondov2015,Roushan2017,chiaro2020direct,Rispoli2019,Lukin2019}. More recently, it has been shown that spatial disorder is not a necessary condition for MBL behavior to emerge, and a gradient magnetic field applied to an otherwise translation-invariant chain of interacting spin-$1/2$ particles is sufficient to induce so-called Stark-MBL \cite{Schulz2019}. From the perspective of the eigenstate thermalization hypothesis (ETH) \cite{Rigol_review,Deutsch_review}, it is not surprising that disorder- and Stark-MBL systems may not thermalize. Indeed, such models are inherently nonergodic, and can be shown to host an extensive number of conserved (quasi)local integrals of motion, similarly to integrable models, and this will prevent thermalization \cite{Abanin_review,gunawardana2022dynamical}. 
Moreover, it has been shown that MBL behavior can still occur in quantum many-body models without any spatial disorder or inhomogeneity. This can happen when the model hosts a local gauge symmetry and the initial state is prepared in a superposition of gauge superselection sectors \cite{Smith2017,Brenes2018}. Localization may then arise even when the model itself is nonintegrable, and, therefore, local observables in the wake of a quench are expected to thermalize. This type of disorder-free localization (DFL) has been shown to exist in various models hosting local symmetries and even in two spatial dimensions \cite{smith2017absence,Metavitsiadis2017,Smith2018,Russomanno2020,Papaefstathiou2020,karpov2021disorder,hart2021logarithmic,Zhu2021,Sous2021}. The main mechanism behind it lies in the initial state being a superposition of an extensive number of gauge superselection sectors, which dynamically induces an effective disorder via the different background charges associated with these sectors. Crucially, this emergent disorder is time-independent, because different gauge superselection sectors do not couple when the dynamics is propagated by the ideal gauge theory, making the background charges fully static.

However, it has been shown that DFL is not stable in the presence of gauge-breaking perturbations \cite{Smith2018}, and thus can only emerge under fine-tuned conditions. This is due to the fact that gauge-breaking errors will couple the different gauge superselection sectors, effectively rendering the emergent disorder over these sectors time-dependent, which eventually destroys localization. Recent works have presented experimentally feasible translation-invariant disorder-free gauge-protection schemes based on quantum Zeno dynamics that have demonstrated successful stabilization of DFL up to times at most quadratic in the protection strength \cite{Halimeh2021stabilizingDFL,Halimeh2021enhancing}.

In this work, we extend these schemes by presenting \textit{Stark gauge protection} (SGP), which involves a linear weighted sum of the local generators or ``pseudogenerators'' \cite{Halimeh2021stabilizing}, where the weights are of a form resembling a Stark potential, hence the nomenclature. We show using exact diagonalization (ED) and Krylov-subspace methods that SGP stabilizes DFL up to all numerically accessible times at relatively small values of the protection strength. We argue that this localized behavior is not due to \textit{bona fide} Stark-MBL, because (i) the gauge symmetry itself is stabilized up to all accessible times, (ii) starting in a translation- and gauge-invariant initial state will in general not lead to localization, and (iii) SGP does not include weak disorder or a harmonic potential, which has been found to be a necessary ingredient for Stark-MBL \cite{Taylor2020}.

The rest of the paper is organized as follows: In Sec.~\ref{sec:SGP}, we review the problem of experimentally relevant gauge-breaking errors and the various schemes to protect against them, before introducing the concept of Stark gauge protection. In Sec.~\ref{sec:results} we present our numerical results for a spin-$S$ $\mathrm{U}(1)$ quantum link model (Sec.~\ref{sec:U1QLM}) and a $\mathbb{Z}_2$ lattice gauge theory (Sec.~\ref{sec:Z2LGT}). We conclude and provide an outlook in Sec.~\ref{sec:conc}. We supplement our work with further numerical results in Appendix~\ref{app:supp}, and we provide the details of our Magnus expansion in Appendix~\ref{app:ME}.

\section{Stark gauge protection}\label{sec:SGP}
Gauge theories are fundamental frameworks for the description of interactions between elementary particles as mediated by gauge bosons \cite{Weinberg_book}. They give rise to an abundance of local constraints that impose an intrinsic relationship between the distribution of charged matter and the resulting electromagnetic field at each point in space and time \cite{Gattringer_book}. These local constraints, such as Gauss's law as a prominent example from quantum electrodynamics, are a manifestation of the gauge symmetry of these models \cite{Zee_book}.

Currently, there are huge experimental efforts towards implementations of gauge theories \cite{Martinez2016,Muschik2017,Bernien2017,Klco2018,Kokail2019,Goerg2019,Schweizer2019,Mil2020,Klco2020,Yang2020,Zhou2021}, given that recent progress in synthetic quantum matter \cite{Bloch2008} has made their quantum simulation a realistic prospect \cite{Wiese_review,Pasquans_review,Alexeev_review,aidelsburger2021cold,zohar2021quantum,klco2021standard,homeier2020mathbbz2}. This has in turn led to a flurry of works on how to stabilize gauge symmetry in realizations of these models on quantum simulators \cite{Zohar2011,Zohar2012,Zohar2013,Hauke2013,Stannigel2014,Kuehn2014,Kuno2015,Yang2016,Kuno2017,Negretti2017,Dutta2017,Barros2019,Halimeh2020a,Lamm2020,Halimeh2020e,Kasper2021nonabelian,Halimeh2021gauge}. Here, we briefly review gauge theories and their stabilization schemes, before introducing the concept of Stark gauge protection.

Let us consider an Abelian quantum gauge theory described by the Hamiltonian $\hat{H}_0$ with local gauge-symmetry generators $\hat{G}_j$ such that $\big[\hat{H}_0,\hat{G}_j\big]=0,\,\forall j$. These commutation relations embody gauge invariance of the model. The symmetry generator $\hat{G}_j$ imposes a local constraint, which in an Abelian gauge theory is usually defined over a matter site and its neighboring gauge links (see Sec.~\ref{sec:results} for details). As such, the subscript $j$ indicates the matter site with which $\hat{G}_j$ is associated. A state $\ket{\phi}$ is said to be gauge-invariant when it satisfies $\hat{G}_j\ket{\phi}=g_j\ket{\phi},\,\forall j$, where the generators' eigenvalues $g_j$ are so-called \textit{background charges}. In a system of $L$ matter sites, the set of background charges $\mathbf{g}=(g_1,g_2,\ldots,g_L)$ defines a unique gauge superselection sector. Since $\big[\hat{H}_0,\hat{G}_j\big]=0,\,\forall j$, the gauge-theory Hamiltonian $\hat{H}_0$ can be block-diagonalized with respect to the gauge superselection sectors, and no coupling between these sectors occurs under the action of $\hat{H}_0$.

In a realistic synthetic quantum matter implementation of gauge theories with both dynamical matter and gauge fields, unavoidable gauge-breaking errors $\lambda\hat{H}_1$ at strength $\lambda$ will arise, where $\big[\hat{H}_1,\hat{H}_0\big]\neq0$ and $\big[\hat{H}_1,\hat{G}_j\big]\neq0$. These errors will break the gauge symmetry generated by $\hat{G}_j$ and couple different gauge superselection sectors. For a system initially prepared in a given \textit{target} gauge sector $\mathbf{g}^\text{tar}=(g_1^\text{tar},g_2^\text{tar},\ldots,g_L^\text{tar})$, several protection schemes have been proposed to stabilize gauge invariance in the presence of errors. One is to make the target sector a ground-state manifold by adding a large penalty term $V\sum_j\big(\hat{G}_j-g_j^\text{tar}\big)^2$ with protection strength $V\gg\lambda$ \cite{Halimeh2020a}. In gauge-theory quench dynamics, this has been shown to stabilize gauge invariance up to all accessible times for finite systems in ED \cite{Halimeh2020a,Halimeh2021gauge}, and for infinite systems in infinite matrix product state (iMPS) calculations \cite{vandamme2021reliability}, which work directly in the thermodynamic limit. This scheme has also been demonstrated to stabilize gauge invariance in equilibrium, leading to a rich \textit{gauge-violation quantum phase diagram} \cite{VanDamme2020}. A disadvantage of this scheme is that it involves terms quadratic in $\hat{G}_j$, which can be experimentally quite challenging to implement.

Another scheme is based on a linear weighted sum of the gauge-symmetry local generators, $V\hat{H}_G=V\sum_jc_j\hat{G}_j$ , which allows for much more experimental feasibility \cite{Halimeh2020e}. Instead of making a ground-state manifold out of the target sector, this \textit{linear gauge protection} energetically isolates it from other sectors. This isolation can be analytically shown to last up to timescales exponential in $V$ when the coefficients $c_j$ are \textit{compliant}, i.e., they are rational numbers satisfying the relation $\sum_jc_j\big(g_j-g_j^\text{tar}\big)=0\iff g_j=g_j^\text{tar},\,\forall j$. However, to guarantee the compliance condition, the sequence $c_j$ will have to grow exponentially with system size $L$, which becomes unfeasible in large-scale quantum simulations of gauge theories. It turns out that a simpler experimentally friendly sequence such as $c_j=(-1)^j$ can stabilize gauge invariance very well based on the quantum Zeno effect (QZE) \cite{facchi2002quantum,facchi2004unification,facchi2009quantum,burgarth2019generalized} in the case of local errors \cite{Halimeh2020e}. For a gauge-invariant initial state, this noncompliant sequence has been numerically shown to stabilize gauge invariance up to all accessible times in both finite \cite{Halimeh2020e} and infinite-size \cite{vandamme2021reliability} spin-$S$ $\mathrm{U}(1)$ quantum link models (QLMs), exceeding analytic predictions based on the QZE. In $\mathbb{Z}_2$ lattice gauge theories (LGTs) where $\hat{G}_j^2=\hat{G}_j$, this scheme has been extended using \textit{local pseudogenerators} (LPGs) $\hat{W}_j$, which act identically to $\hat{G}_j$ in the target sector but not necessarily outside of it \cite{Halimeh2021stabilizing}. In particular, they satisfy the relation $\hat{W}_j\ket{\phi}=g_j^\text{tar}\ket{\phi}\iff \hat{G}_j\ket{\phi}=g_j^\text{tar}\ket{\phi}$ at each local constraint associated with matter site $j$. The linear protection term then takes the form $V\hat{H}_W=V\sum_jc_j\big(\hat{W}_j-g_j^\text{tar}\big)$, and the conclusions from Ref.~\cite{Halimeh2020e} apply the same way, with stabilization of gauge invariance having been numerically demonstrated up to all accessible times in finite systems \cite{Halimeh2021stabilizing} and also recently in the thermodynamic limit \cite{vandamme2021suppressing}, even when $c_j$ is a noncompliant repeating sequence over two or four matter sites.

If the system is initially prepared in a superposition of an extensive number of gauge superselection sectors, the effective disorder over their associated background charges will become time-dependent in the presence of gauge-breaking errors, thereby destroying DFL and leading to thermalization \cite{Smith2018}. Since fine-tuning is very difficult to achieve in an experiment, stabilizing DFL becomes crucial. Even though the linear protection schemes of Refs.~\cite{Halimeh2020e,Halimeh2021stabilizing} have originally been devised for quenches starting in a target gauge sector, they have surprisingly performed well for quenches starting in a superposition over an extensive number of gauge sectors. Indeed, using translation-invariant sequences $c_j$ that alternate between odd and even sites, linear gauge protection schemes have stabilized DFL \cite{Halimeh2021stabilizingDFL,Halimeh2021enhancing} up to times at least linear in the protection strength in agreement with the worst-case prediction of the QZE, and at best up to timescales quadratic in the protection strength. Naturally, it is of great interest to devise a disorder-free stabilization scheme that can restore DFL up to all accessible evolution times.

In this paper, we achieve the latter through employing the disorder-free Stark gauge protection, which for the spin-$S$ $\mathrm{U}(1)$ QLM takes the form
\begin{align}\nonumber
    V\hat{H}_\text{SGP}&=V\sum_jj(-1)^j\hat{G}_j\\\label{eq:U1QLM_SGP}
    &=V\sum_jj\big(\hat{n}_j+\hat{s}^z_{j-1,j}+\hat{s}^z_{j,j+1}\big).
\end{align}
In Sec.~\ref{sec:U1QLM}, we will focus on the case of the $\mathrm{U}(1)$ QLM and show how SGP stabilizes DFL up to all accessible times in our numerical results. In Sec.~\ref{sec:Z2LGT}, we will show how SGP can be extended to the case of the $\mathbb{Z}_2$ LGT by utilizing the associated LPG \cite{Halimeh2021stabilizing}. We show the robustness of our conclusions with respect to system size, where our results suggest better performance of SGP with increasing system size. As we show analytically through a Magnus expansion (see details in Appendix~\ref{app:ME}), the addition of the SGP term at moderate or large protection strength $V$ leads to the emergence of an effective Hamiltonian where gauge-breaking terms are not only suppressed by $V$, but also by the matter-site index appearing in the Stark potential. As we derive in Appendix~\ref{app:ME}, the gauge errors are locally suppressed $\propto1/[(2j+1)V]$. We argue that this is the crucial ingredient in stabilizing DFL up to all accessible evolution times.

\section{Results and discussion}\label{sec:results}
We now present our main numerical results obtained from ED and Krylov-based time evolution \cite{Moler2003,EXPOKIT} for the quench dynamics under $\hat{H}=\hat{H}_0+\lambda\hat{H}_1+V\hat{H}_\text{pro}$, where $V\hat{H}_\text{pro}$ is one of several protection schemes employed in this work, including $V\hat{H}_\text{SGP}$. We will compare the performance of SGP to other protection schemes both numerically and analytically.

\subsection{$\mathrm{U}(1)$ quantum link model}\label{sec:U1QLM}
We first consider the paradigmatic $(1+1)-$dimensional spin-$S$ $\mathrm{U}(1)$ QLM given by the Hamiltonian \cite{Wiese_review,Chandrasekharan1997,Hauke2013,Yang2016,Kasper2017}
\begin{align}\nonumber
    \hat{H}_0=\sum_{j=1}^L\bigg[&\frac{J}{2a\sqrt{S(S+1)}}\big(\hat{\sigma}^-_j\hat{s}^+_{j,j+1}\hat{\sigma}^-_{j+1}+\text{H.c.}\big)\\\label{eq:H0}
    &+\frac{\mu}{2}\hat{\sigma}^z_j+\frac{\kappa^2a}{2}\big(\hat{s}^z_{j,j+1}\big)^2\bigg].
\end{align}
The first term of this Hamiltonian describes the creation and annihilation of matter along with the concomitant change in the electric field in order to preserve the gauge symmetry. The Pauli ladder operators $\hat{\sigma}^{\pm}_j$ represent the creation and annihilation of matter at lattice site $j$, where the matter occupation there is given by $\hat{n}_j=\big(\hat{\sigma}^z_j+\mathds{1}\big)/2$ with mass $\mu$. The number of matter sites is denoted by $L$. The spin-$S$ ladder operators $\hat{s}^{\pm}_{j,j+1}$ describe the action of the gauge field at the link between matter sites $j$ and $j+1$, and $\hat{s}^z_{j,j+1}$ is the electric field on that link, with coupling strength $\kappa$. The lattice spacing $a$ is set to unity throughout this work. In the limits of $a\to0$ and $S\to\infty$, Eq.~\eqref{eq:H0} retrieves the quantum field theory limit of quantum electrodynamics, although convergence to that limit can be achieved already with finite values of $a$ and small values of $S$ in and out of equilibrium \cite{Buyens2017,Banuls2018,Banuls2020,Zache2021achieving,Halimeh2021achieving}. Throughout our paper, we will set the energy scale to $J=1$.

The $\mathrm{U}(1)$ gauge symmetry of Eq.~\eqref{eq:H0} is generated by the operator
\begin{align}\label{eq:U1QLM_Gj}
    \hat{G}_j=(-1)^j\big(\hat{n}_j+\hat{s}^z_{j-1,j}+\hat{s}^z_{j,j+1}\big),
\end{align}
whose eigenvalues $g_j$ are the background charges, and where $\big[\hat{H}_0,\hat{G}_j\big]=0,\,\forall j$. A set of eigenvalues $g_j$ over the lattice define a gauge superselection sector $\mathbf{g}=\big(g_1,g_2,\ldots,g_L\big)$.

\begin{figure}[t!]
	\centering
	\includegraphics[width=.4\textwidth]{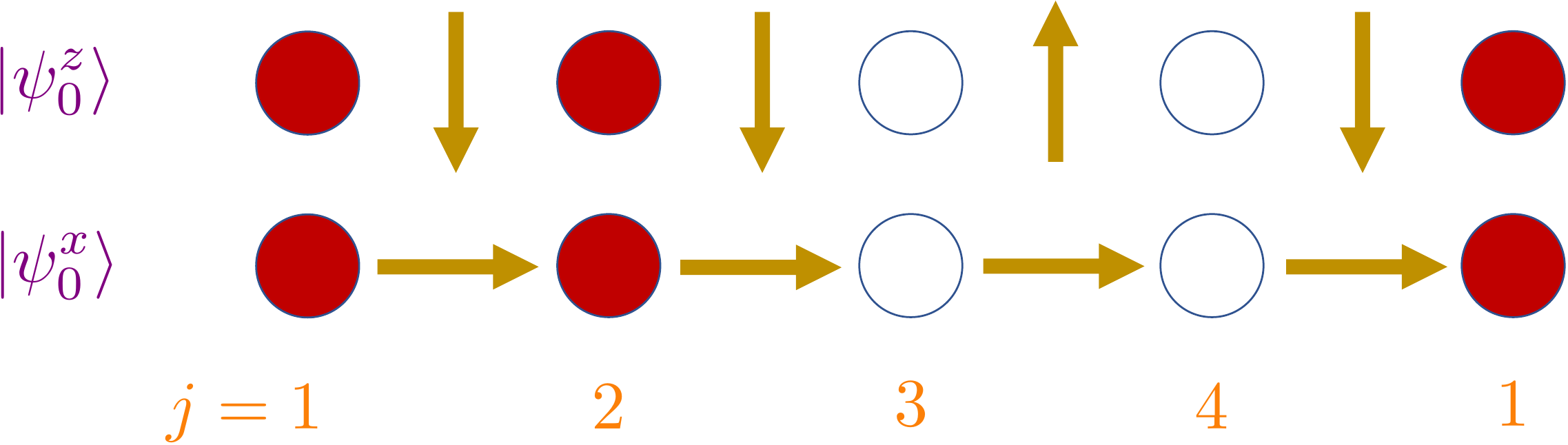}
	\caption{(Color online). Domain-wall initial states used in the case of the $\mathrm{U}(1)$ quantum link model. The gauge-invariant state $\ket{\psi_0^z}$ resides in the gauge superselection sector $g_j=0,\,\forall j$. The initial state $\ket{\psi_0^x}$ is not gauge-invariant, and forms a superposition over an extensive number of gauge superselection sectors.}
	\label{fig:U1QLM_InitialStates} 
\end{figure}

When the system is prepared in a translation-invariant gauge-symmetric initial state, such as $\ket{\psi_0^z}$ in Fig.~\ref{fig:U1QLM_InitialStates}, it is expected to thermalize in the long-time limit after a quench by $\hat{H}_0$, because the Hamiltonian~\eqref{eq:H0} is ergodic in the gauge superselection sector $g_j=0,\,\forall j$. Counterintuitively, when the initial state is a translation-invariant superposition of an extensive number of gauge superselection sectors, such as $\ket{\psi_0^x}$ in Fig.~\ref{fig:U1QLM_InitialStates}, the dynamics localizes up to all times in the wake of a quench under the nonintegrable Hamiltonian $\hat{H}_0$, and the system permanently retains memory of its initial state. This localization is due to the superposition dynamically inducing an effective disorder over the background charges of the different gauge superselection sectors. Intriguingly, this leads to DFL even when the model~\eqref{eq:H0} is disorder-free and nonintegrable, and there is no disorder in the initial state $\ket{\psi^x_0}$ \cite{Smith2017,Brenes2018}.

However, it has been shown that DFL requires fine-tuning, and even perturbative gauge-breaking errors can destroy it \cite{Smith2018}. An example of such errors that may arise in a realistic implementation \cite{Mil2020} of the spin-$S$ $\mathrm{U}(1)$ QLM takes the form
\begin{align}\label{eq:H1}
    \lambda\hat{H}_1=\lambda\sum_j\Bigg[\hat{\sigma}^-_j\hat{\sigma}^-_{j+1}+\frac{\hat{s}^+_{j,j+1}}{2\sqrt{S(S+1)}}+\text{H.c.}\Bigg].
\end{align}
The Hamiltonian $\hat{H}_1$ describes the creation and annihilation of matter without a corresponding simultaneous action from the gauge field and vice versa, leading to the violation of Gauss's law, and hence, the destruction of DFL.

Recently, a disorder-free single-body protection scheme has been proposed,
\begin{align}
    V\hat{H}_G=V\sum_j(-1)^j\hat{G}_j,
\end{align}
which stabilizes DFL up to timescales quadratic in the protection strength $V$ \cite{Halimeh2021stabilizingDFL}. The underlying principle is the quantum Zeno dynamics \cite{facchi2002quantum,facchi2004unification,facchi2009quantum,burgarth2019generalized}, which guarantees a controlled gauge violation up to timescales at least linear in $V$. After these timescales, DFL quickly vanishes giving way to thermalization \cite{Halimeh2021stabilizingDFL}.

\begin{figure}[t!]
	\centering
	\includegraphics[width=.48\textwidth]{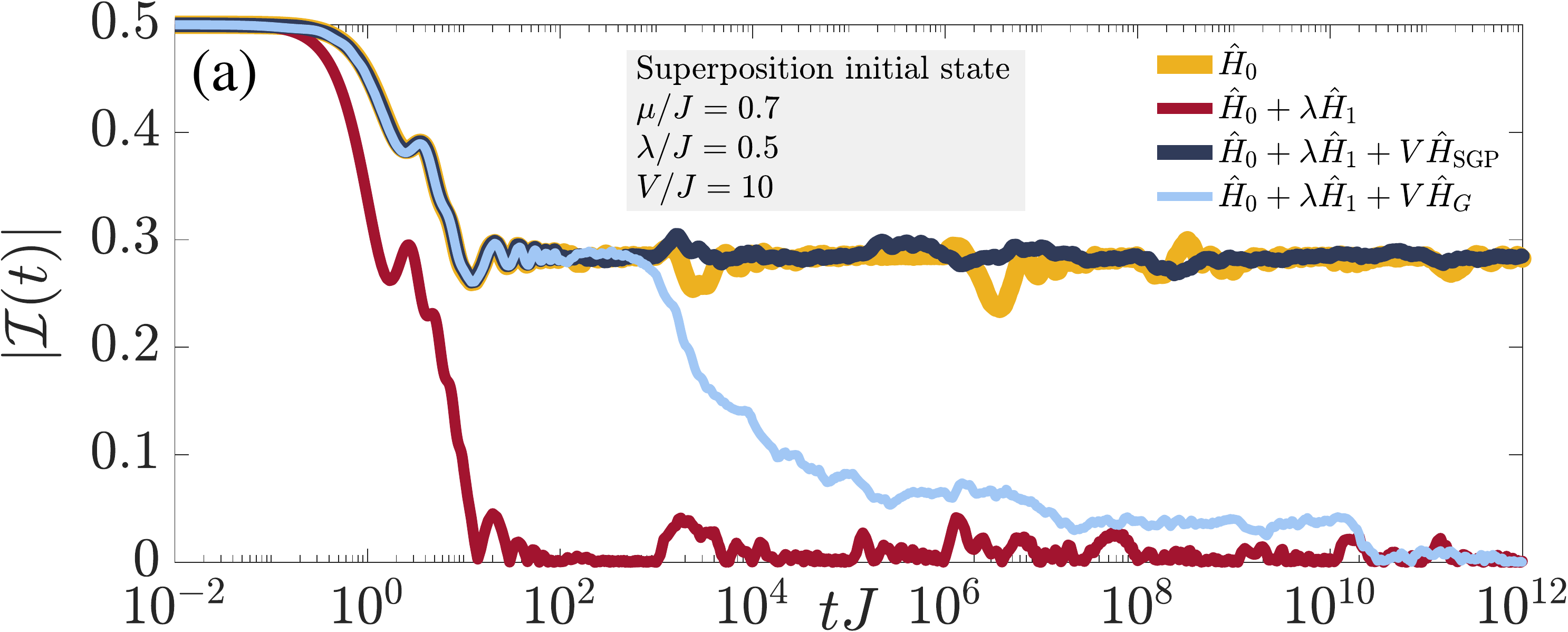}\\
	\vspace{2.1mm}
	\includegraphics[width=.48\textwidth]{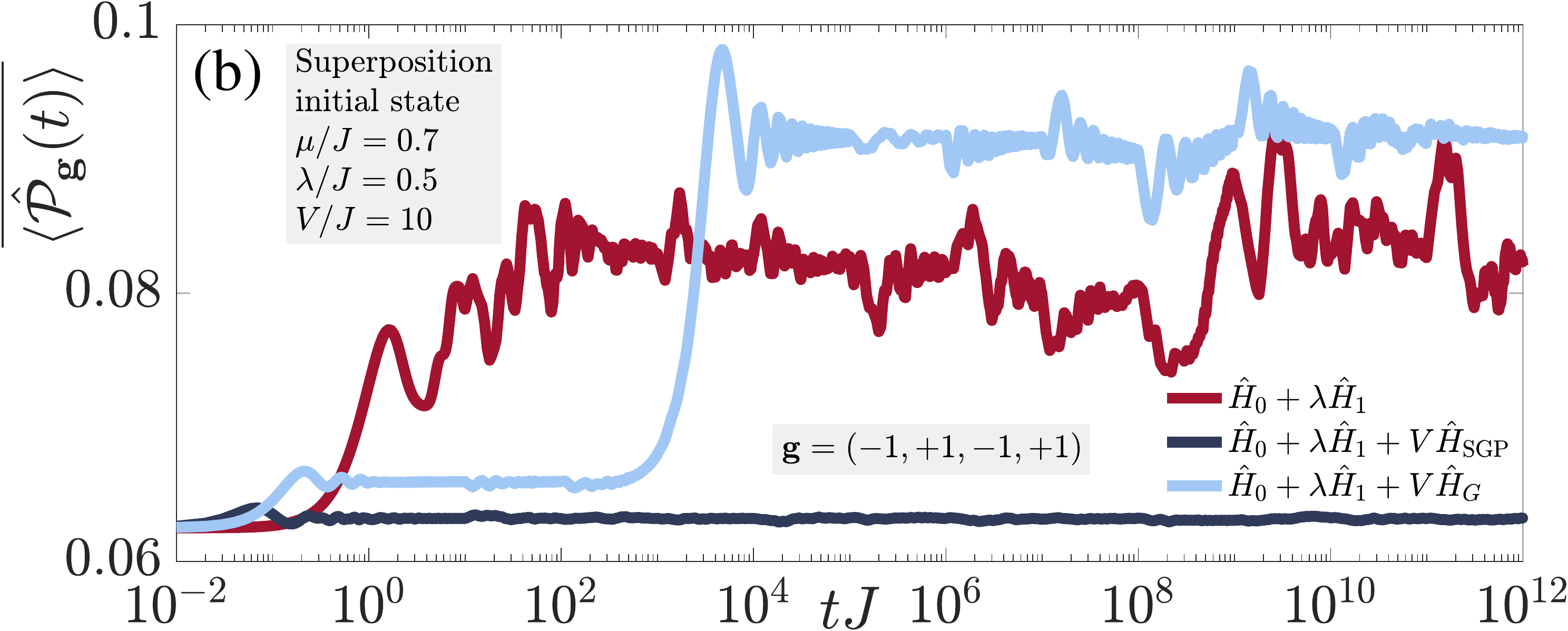}\\
	\vspace{2.1mm}
	\includegraphics[width=.48\textwidth]{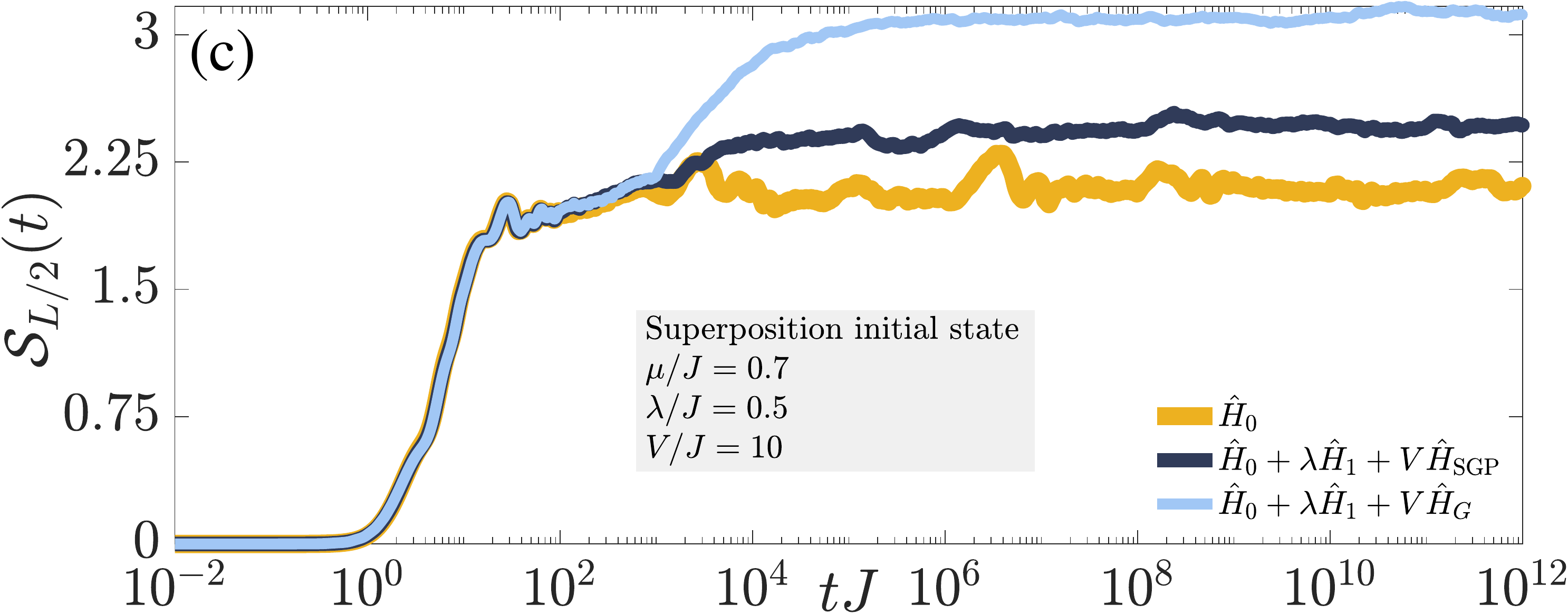}
	\caption{(Color online). Stark gauge protection in the spin-$1/2$ $\mathrm{U}(1)$ quantum link model. The system is initialized in the domain-wall state $\ket{\psi_0^x}$ shown in Fig.~\ref{fig:U1QLM_InitialStates}, which is a superposition over an extensive number of gauge superselection sectors, and then quenched by $\hat{H}=\hat{H}_0+\lambda\hat{H}_1+V\hat{H}_\text{pro}$. Results are computed in exact diagonalization for $L=4$ matter sites and $L=4$ gauge links with periodic boundary conditions. (a) Disorder-free localization persists for all accessible evolution times in the ideal case, where the imbalance~\eqref{eq:imbalance} does not thermalize to zero but rather settles into a plateau with nonzero value for all times (yellow curve, $\lambda=V=0$). However, unavoidable errors ($\lambda\neq0$) destroy DFL in the absence of protection (red curve, $V=0$). When employing the staggered protection $V\hat{H}_\text{pro}=V\hat{H}_G=V\sum_j(-1)^j\hat{G}_j$, DFL is restored up to a timescale $\propto V^2/(J\lambda^2)$ (light blue curve) \cite{Halimeh2021stabilizingDFL}. In this work, we introduce the concept of Stark gauge protection $V\hat{H}_\text{pro}=V\hat{H}_\text{SGP}=V\sum_jj(-1)^j\hat{G}_j$, which for the same value of protection strength $V=10J$ restores DFL up to all accessible evolution times. (b) The superior performance of SGP is also evident in the expectation values of the projectors $\hat{\mathcal{P}}_\mathbf{g}$ onto the different superselection sectors $\mathbf{g}$, where it stabilizes $\langle\hat{\mathcal{P}}_\mathbf{g}\rangle$ near its initial value up to all accessible evolution times, while the staggered protection stabilizes it only up to times $\propto V^2/(J\lambda^2)$. (c) The mid-chain entanglement entropy also shows better long-time localization under SGP protection than its staggered counterpart.}
	\label{fig:U1QLM_comparison} 
\end{figure}

We now compare this scheme to that of SGP~\eqref{eq:U1QLM_SGP} by first looking at the imbalance
\begin{align}\label{eq:imbalance}
    \mathcal{I}(t)=\frac{1}{Lt}\int_0^tds\sum_{j=1}^Lp_j\bra{\psi(s)}\hat{n}_j\ket{\psi(s)},
\end{align}
where $\ket{\psi(t)}=e^{-i\hat{H}t}\ket{\psi_0}$, $\hat{H}=\hat{H}_0+\lambda\hat{H}_1+V\hat{H}_\text{pro}$ is the \textit{faulty} gauge theory that we attempt to stabilize with the protection term $V\hat{H}_\text{pro}$, $\ket{\psi_0}$ is the initial state, and $p_j=\bra{\psi_0}\hat{\sigma}^z_j\ket{\psi_0}$. Let us first initialize our system in the domain-wall initial state $\ket{\psi_0^x}$ shown in Fig.~\ref{fig:U1QLM_InitialStates}, which is a superposition over an extensive number of gauge superselection sectors due to $\hat{G}_j$ defined in Eq.~\eqref{eq:U1QLM_Gj}. The corresponding quench dynamics is shown in Fig.~\ref{fig:U1QLM_comparison}, calculated in ED for $L=4$ matter sites and $L=4$ gauge links with periodic boundary conditions. As we will show later, our conclusions also hold for larger system sizes.

Quenching with the ideal gauge theory $\hat{H}_0$, the imbalance~\eqref{eq:imbalance} is expected to decay to zero for a translation- and gauge-invariant initial state such as $\ket{\psi_0^z}$ due to thermalization, and indeed it does \cite{Halimeh2021stabilizingDFL}. On the other hand, for the superposition initial state $\ket{\psi_0^x}$, the system will permanently retain memory of its initial condition and, as a result, the imbalance will settle into a nonzero plateau that persists over all accessible evolution times, as shown in Fig.~\ref{fig:U1QLM_comparison}(a). However, in realistic experimental implementations of gauge theories with dynamical matter and gauge fields, there will always be unavoidable errors $\lambda\hat{H}_1$ at strength $\lambda\neq0$ that will undermine gauge invariance. The effect of these errors can be drastic, destroying DFL fairly quickly \cite{Smith2018,Halimeh2021stabilizingDFL}, as demonstrated by the red curve in Fig.~\ref{fig:U1QLM_comparison}(a) where $\lambda=0.5J$. Upon adding the SGP term~\eqref{eq:U1QLM_SGP} at an experimentally friendly value of the protection strength $V=10J$, the DFL plateau is restored qualitatively up to all accessible evolution times in ED. This is contrasted with the staggered protection in the form of $V\hat{H}_G=V\sum_j(-1)^j\hat{G}_j$, which stabilizes DFL only up to a timescale $\propto V^2/(J\lambda^2)$ \cite{Halimeh2021stabilizingDFL}.

This picture is corroborated by looking at the expectation value of the projector $\hat{\mathcal{P}}_\mathbf{g}$ onto the gauge superselection sector $\mathbf{g}=(-1,+1,-1,+1)$ in Fig.~\ref{fig:U1QLM_comparison}(b). In the dynamics under the ideal theory, $\langle\hat{\mathcal{P}}_\mathbf{g}\rangle$ will always remain constant, but in the presence of unprotected errors, it deviates immediately from its initial value and exhibits more or less violent dynamics. Whereas the staggered protection stabilizes $\langle\hat{\mathcal{P}}_\mathbf{g}\rangle$ at a plateau different from its initial value up to a timescale $\propto V^2/(J\lambda^2)$ after which it diverges quickly, the SGP stabilizes it very close to its initial value up to all accessible evolution times in ED. This strongly suggests that the localization restored due to the SGP is inherently related to the gauge symmetry of the model. We have checked that our conclusions are qualitatively the same when considering projectors onto gauge superselection sectors other than $\mathbf{g}=(-1,+1,-1,+1)$.

We now look at the mid-chain entanglement entropy $\mathcal{S}_{L/2}(t)$ in Fig.~\ref{fig:U1QLM_comparison}(c). Both the dynamics of $\mathcal{S}_{L/2}(t)$ under the staggered protection and SGP exactly reproduce the ideal case up to a given time, after which the dynamics with staggered protection shows $\mathcal{S}_{L/2}(t)$ growing significantly larger than in the case of SGP. This is indicative of stabilized localization up to all times under SGP, in contrast to staggered protection.

It is to be noted that even though we have used $\mu=0.7J$ in our ED calculations of Fig.~\ref{fig:U1QLM_comparison}, we have checked that our conclusions are valid for different values of $\mu$. Furthermore, our conclusions also lend themselves to larger link spin lengths $S>1/2$ (see Appendix~\ref{app:supp}).

\begin{figure}[t!]
	\centering
	\includegraphics[width=.48\textwidth]{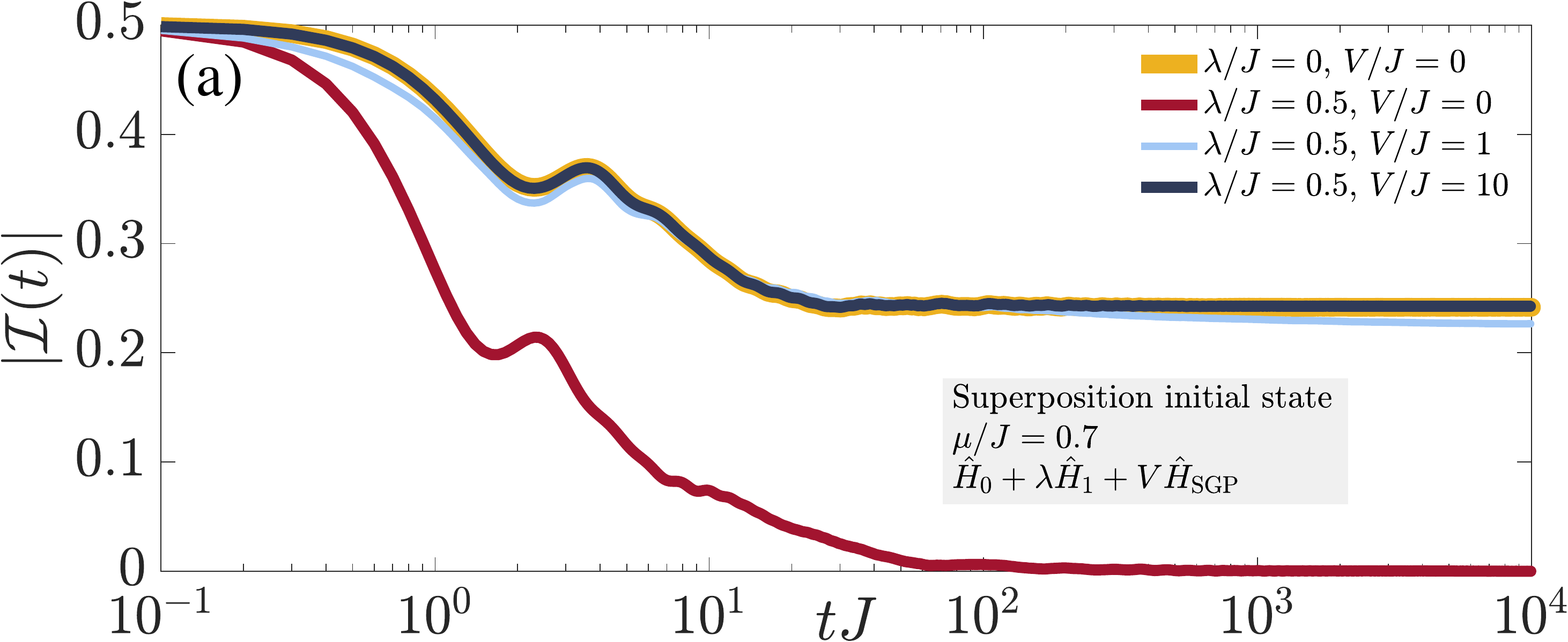}\\
	\vspace{2.1mm}
	\includegraphics[width=.48\textwidth]{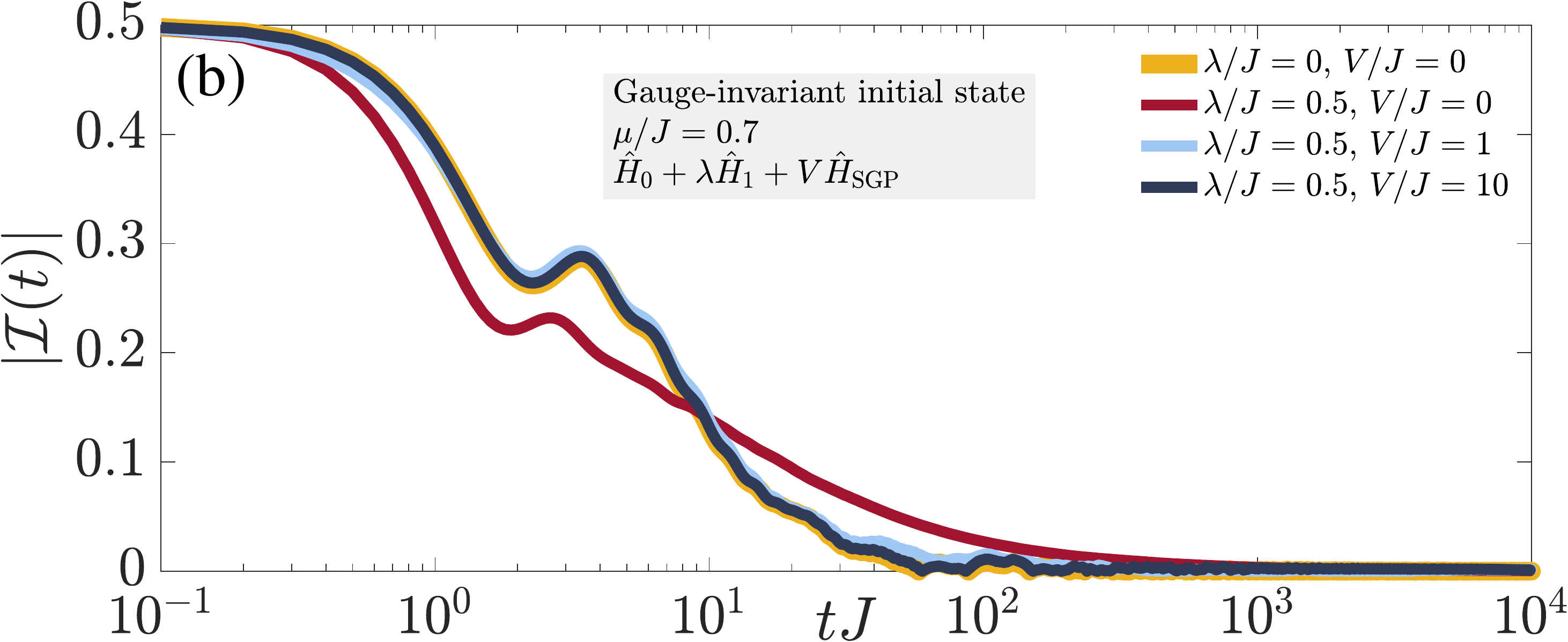}
	\caption{(Color online). Quench dynamics under the faulty theory $\hat{H}_0+\lambda\hat{H}_1+V\hat{H}_\text{SGP}$, where $\hat{H}_0$ is the Hamiltonian~\eqref{eq:H0} of the $\mathrm{U}(1)$ QLM with $S=1/2$, and $\hat{H}_\text{SGP}$ is the corresponding Stark gauge protection~\eqref{eq:U1QLM_SGP}. The error term is given in Eq.~\eqref{eq:H1}. These numerical results are obtained using Krylov-based methods for chains of $L=8$ matter sites and $L=8$ gauge links with periodic boundary conditions. (a) Starting in a domain-wall initial state that is a superposition over an extensive number of gauge sectors, such as $\ket{\psi_0^x}$ in Fig.~\ref{fig:U1QLM_InitialStates}, the system does not thermalize in the ideal case ($\lambda=V=0$), and disorder-free localization is prominent up to all accessible evolution times. In the presence of errors without any protection, DFL is destroyed, and the system thermalizes with the imbalance going to zero rather quickly. Upon adding SGP, already small values of the protection strength such as $V=J$ stabilize DFL quite well. At $V=10J$, we find that SGP restores DFL not just qualitatively but also quantitatively for all accessible evolution times, outperforming the same case for the smaller system size shown in Fig.~\ref{fig:U1QLM_comparison}(a). (b) Starting in a gauge-invariant domain-wall initial state such as $\ket{\psi_0^z}$ of Fig.~\ref{fig:U1QLM_InitialStates}, we see that the imbalance vanishes in all cases. When SGP is turned on, it merely quantitatively restores the ideal thermalizing dynamics in this case, but no localization is present.}
	\label{fig:U1QLM_Krylov} 
\end{figure}

We have employed a Magnus expansion in order to arrive at an effective Hamiltonian describing the dynamics of the faulty gauge theory with staggered protection or SGP. As detailed in Appendix~\ref{app:ME}, the effective Hamiltonian differs drastically depending on the type of protection. Whereas for the staggered protection the gauge-breaking terms in the effective Hamiltonian are suppressed only by $V$, in the case of SGP they are locally suppressed by $(2j+1)V$; see Appendix~\ref{app:ME}. This naturally leads to better stabilization of gauge invariance under SGP, and indicates that this protection scheme may maintain its performance in the thermodynamic limit. 

In order to check this, we have employed a Krylov-based time-evolution method in order to calculate the dynamics of the imbalance~\eqref{eq:imbalance} for the spin-$1/2$ $\mathrm{U}(1)$ QLM with $L=8$ matter sites and $L=8$ gauge links under periodic boundary conditions, as shown in Fig.~\ref{fig:U1QLM_Krylov}. When the initial state is a domain wall in the matter fields with the electric fields locally aligned such that the wave function is a superposition over an extensive number of superselection sectors (an extension of $\ket{\psi_0^x}$ of Fig.~\ref{fig:U1QLM_InitialStates} to $L=8$ matter sites and $L=8$ gauge links), we find that DFL is more or less restored already at $V=J$ in Fig.~\ref{fig:U1QLM_Krylov}(a). In fact, for $V=10J$, the DFL plateau of the ideal case is exactly reproduced quantitatively for all accessible evolution times. Quantitatively, this is better performance than the case of $L=4$ matter sites shown in Fig.~\ref{fig:U1QLM_comparison}(a). There, the DFL plateau is restored qualitatively for all times, but agrees quantitatively with that of the ideal case up to $t\approx10^2/J$. On the other hand, for the larger system size shown in Fig.~\ref{fig:U1QLM_Krylov}(a), the ideal DFL plateau is restored up to at least two orders of magnitude longer, i.e., $t=10^4/J$. This is in agreement with our expectation that with larger system sizes gauge-breaking errors will be further suppressed when employing SGP as this suppression is locally proportional to the matter-site index $j$. This full quantitative agreement between the dynamics under the faulty theory with SGP and under the ideal case indicates that the DFL appearing in the presence of SGP is due to the $\mathrm{U}(1)$ gauge symmetry of the ideal theory, and not due to Stark-MBL.

In order to further rule out that the localization seen in Figs.~\ref{fig:U1QLM_comparison} and~\ref{fig:U1QLM_Krylov}(a) is due to Stark-MBL, we consider the gauge-invariant state $\ket{\psi_0^z}$ shown in Fig.~\ref{fig:U1QLM_InitialStates}, albeit for $L=8$ matter sites. The dynamics of its imbalance~\eqref{eq:imbalance} is shown in the wake of a quench with $\hat{H}_0+\lambda\hat{H}_1+V\hat{H}_\text{SGP}$ in Fig.~\ref{fig:U1QLM_Krylov}(b). As expected, the imbalance under the ideal theory ($\lambda=V=0$) vanishes at long times, indicating thermalization (yellow curve). When errors are added without protection, the imbalance also decays to zero, but is qualitatively different from the ideal case in its dynamics (red curve). Upon adding SGP, we see that at any considered value of $V$ the dynamics under the ideal theory is quantitatively retrieved, with the imbalance decaying to zero at long times. This shows that adding SGP merely restores the original gauge symmetry of the ideal $\mathrm{U}(1)$ QLM, and DFL can arise only when the initial state is itself a superposition of an extensive number of gauge superselection sectors. Even though the gauge-invariant initial state used in Fig.~\ref{fig:U1QLM_Krylov}(b) is in the superselection sector $g_j=0,\,\forall j$, we have checked that starting in a translation-invariant initial state in the target sector $g_j=(-1)^j$, for example, also leads to the same conclusion.

\subsection{$\mathbb{Z}_2$ lattice gauge theory}\label{sec:Z2LGT}
We now explore the potential of SGP in implementations of the $\mathbb{Z}_2$ LGT given by the Hamiltonian \cite{Zohar2017,Borla2019,Yang2020fragmentation,kebric2021confinement}
\begin{align}\label{eq:Z2LGT}
    \hat{H}_0=\sum_{j=1}^L\Big[J\big(\hat{a}_j^\dagger\hat{\tau}^z_{j,j+1}\hat{a}_{j+1}+\text{H.c.}\big)-h\hat{\tau}^x_{j,j+1}\Big].
\end{align}
The hard-core bosonic ladder operators $\hat{a}_j^{(\dagger)}$ represent the creation (annihilation) of matter at site $j$, with $\hat{n}_j=\hat{a}_j^\dagger\hat{a}_j$ denoting the hard-core boson number operator there, and the Pauli matrix $\hat{\tau}^{x(z)}_{j,j+1}$ represents the electric (gauge) field at the link between matter sites $j$ and $j+1$. The energy scale is set by $J=1$. The generator of the $\mathbb{Z}_2$ gauge symmetry of this model is
\begin{align}\label{eq:Z2LGT_Gj}
    \hat{G}_j=(-1)^{\hat{n}_j}\hat{\tau}^x_{j-1,j}\hat{\tau}^x_{j,j+1},
\end{align}
and gauge invariance becomes manifest in the commutation relations $\big[\hat{H}_0,\hat{G}_j\big]=0,\,\forall j$. In addition to  the $\mathbb{Z}_2$ gauge symmetry, the $\mathbb{Z}_2$ LGT~\eqref{eq:Z2LGT} hosts a global $\mathrm{U}(1)$ symmetry in the form of boson-number conservation.

Recently, a building block of this model has been implemented in an ultracold atom experiment~\cite{Schweizer2019}, and a superconducting qubit setup has been proposed \cite{homeier2020mathbbz2}. Experimentally relevant errors inspired from these setups take the following form for an extended system:
\begin{align}\nonumber
	\lambda\hat{H}_1=\,\lambda\sum_{j=1}^{L}\Big\{&\Big[\hat{a}_j^\dagger\hat{a}_{j+1}\big(\alpha_1\hat{\tau}^+_{j,j+1} +\alpha_2\hat{\tau}^-_{j,j+1}+1\big)+\mathrm{H.c.}\Big]\\\label{eq:Z2LGT_H1}
	&+\big(\alpha_3\hat{n}_j-\alpha_4\hat{n}_{j+1}+1\big)\hat{\tau}^z_{j,j+1}\Big\},
\end{align}
where the coefficients $\alpha_n$ depend on a dimensionless driving parameter in the experiment of Ref.~\cite{Schweizer2019}. When the latter is set to its optimal value, this renders the coefficients as $\alpha_1=0.5110$, $\alpha_2=-0.4953$, $\alpha_3=0.7696$, and $\alpha_4=0.2147$, where we have normalized them such that their sum is unity. These are the values that we will use for the main results of our work. However, we have checked that different values of $\alpha_n$ do not affect the conclusions of our study. The error term~\eqref{eq:Z2LGT_H1} violates the $\mathbb{Z}_2$ gauge symmetry of the model in Eq.~\eqref{eq:Z2LGT}, $\big[\hat{H}_1,\hat{G}_j\big]\neq0$, but it does conserve boson number, $\big[\hat{H}_1,\sum_j\hat{n}_j\big]=0$, leaving the global $\mathrm{U}(1)$ symmetry of Eq.~\eqref{eq:Z2LGT} intact.

Unlike the case of the $\mathrm{U}(1)$ QLM considered in Sec.~\ref{sec:U1QLM}, gauge protection terms linear in $\hat{G}_j$~\eqref{eq:Z2LGT_Gj} are experimentally very difficult to realize in the case of the $\mathbb{Z}_2$ LGT, since Eq.~\eqref{eq:Z2LGT_Gj} includes three-body terms. A convenient solution is employing gauge protection linear in the pseudogenerator (LPG) \cite{Halimeh2021stabilizing}
\begin{align}\label{eq:LPG}
\hat{W}_j(g^\text{tar}_j)=\hat{\tau}^x_{j-1,j}\hat{\tau}^x_{j,j+1}+2g^\text{tar}_j\hat{n}_j,
\end{align}
where $g^\text{tar}_j$ is the local charge of the target superselection sector $\mathbf{g}^\text{tar}=(g^\text{tar}_1,g^\text{tar}_2,\ldots,g^\text{tar}_L)$. The LPG acts identically to $\hat{G}_j$ in the target sector, satisfying the relation
\begin{align}
    \hat{W}_j\ket{\phi}=g^\text{tar}_j\ket{\phi}\iff\hat{G}_j\ket{\phi}=g^\text{tar}_j\ket{\phi}.
\end{align}

One can then extend the principles of linear gauge protection \cite{Halimeh2020e} by adding the term
\begin{align}\label{eq:LPGpro}
    V\hat{H}_W=V\sum_jc_j\big[\hat{W}_j(g^\text{tar}_j)-g^\text{tar}_j\big],
\end{align}
which stabilizes gauge invariance and penalizes processes away from the target sector. Even more, the term~\eqref{eq:LPGpro} has been shown to also stabilize a gauge theory also when the initial state is a superposition of an extensive number of gauge superselection sectors, protecting and even enhancing DFL in the $\mathbb{Z}_2$ LGT for an alternating sequence $c_j$ up to times linear in the protection strength $V$ based on the QZE \cite{Halimeh2021enhancing}. This enhancement is due to the fact that the local symmetry associated with $\hat{W}_j$ constitutes an extension of the original $\mathbb{Z}_2$ gauge symmetry, which it contains. This therefore leads to an increase in the background charges associated with the superposition initial state, since in the large-$V$ limit the effective Hamiltonian, which can be derived using the formalism of quantum Zeno subspaces, hosts both the symmetry due to $\hat{W}_j$ and the $\mathbb{Z}_2$ gauge symmetry generated by $\hat{G}_j$~\eqref{eq:Z2LGT_Gj}. This leads to a greater effective disorder over more gauge sectors, which in turn enhances DFL \cite{Halimeh2021enhancing}.

\begin{figure}[t!]
	\centering
	\includegraphics[width=.4\textwidth]{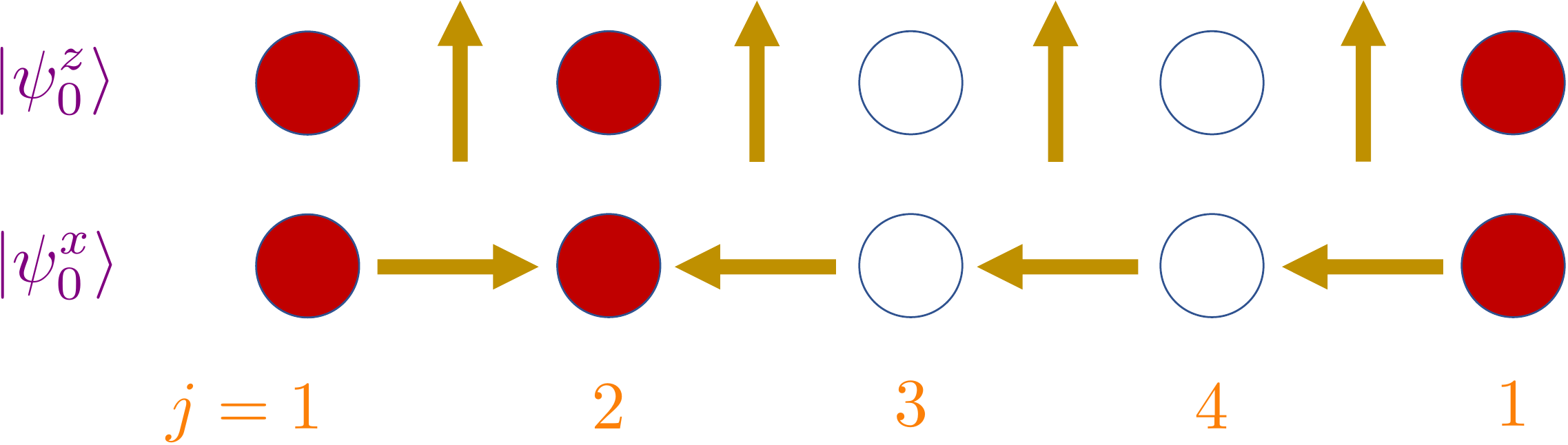}
	\caption{(Color online). Domain-wall initial states used in the case of the $\mathbb{Z}_2$ lattice gauge theory. The gauge-invariant state $\ket{\psi_0^x}$ resides in the gauge superselection sector $g_j=+1,\,\forall j$. The initial state $\ket{\psi_0^z}$ is not gauge-invariant, and forms a superposition over an extensive number of gauge superselection sectors.}
	\label{fig:Z2LGT_InitialStates} 
\end{figure}

\begin{figure}[t!]
	\centering
	\includegraphics[width=.48\textwidth]{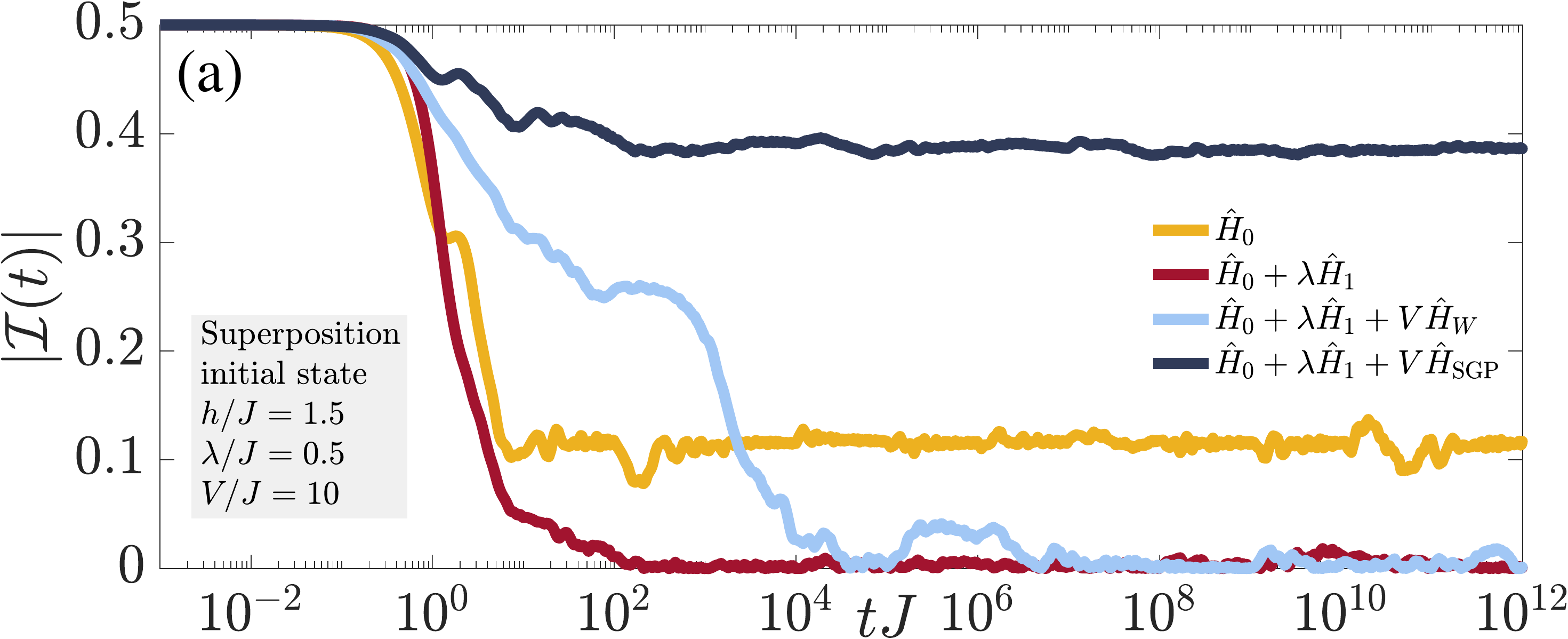}\\
	\vspace{2.1mm}
	\includegraphics[width=.48\textwidth]{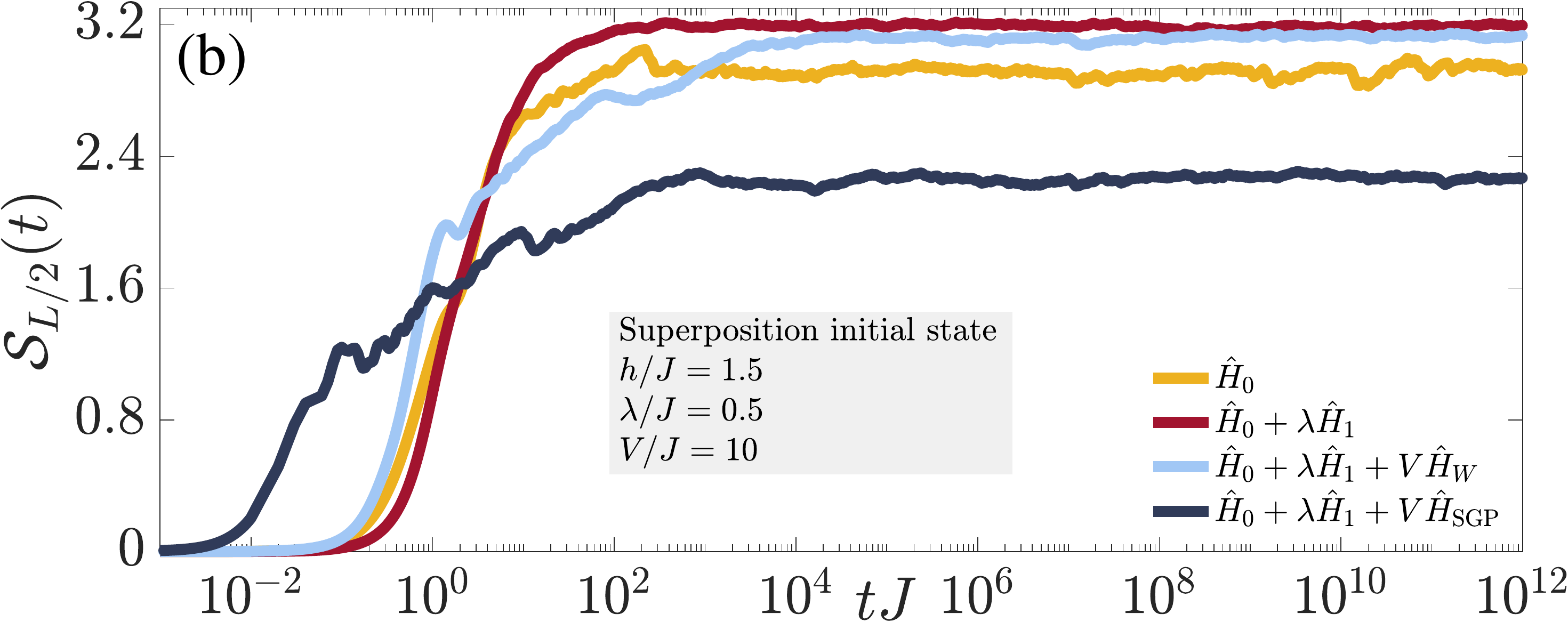}
	\caption{(Color online). Stark gauge protection in the $\mathbb{Z}_2$ lattice gauge theory in the presence of unavoidable experimental errors. The initial state $\ket{\psi_0^z}$ is a superposition of an extensive number of superselection sectors, and a domain wall in the matter fields, such that the left half is occupied while the right half is empty (see Fig.~\ref{fig:Z2LGT_InitialStates}). The results are calculated in exact diagonalization for $L=4$ matter sites and $L=4$ gauge links with periodic boundary conditions. (a) The imbalance~\eqref{eq:imbalance} does not decay to zero in the ideal case, giving rise to a disorder-free localization. In the presence of gauge-breaking errors, DFL is quickly destroyed. Previous works have introduced ``staggered protection'' terms consisting of a linear alternating sum of the local pseudogenerator (staggered protection), a simplification of the gauge-symmetry generator, but that is experimentally simpler to realize and is associated with a richer local symmetry \cite{Halimeh2021stabilizing}. At a moderate protection strength $V=10J$, we see that this scheme does not fare well for an error strength of $\lambda=0.5J$. In contrast, Stark gauge protection~\eqref{eq:Z2LGT_SGP} shows great stabilization of the DFL plateau, now enhanced with respect to the ideal case due to the richer local symmetry associated with the local pseudogenerator (see text for details). (b) This picture is further affirmed in the behavior of the mid-chain entanglement entropy. Whereas the staggered protection protects localization at relatively short times but then delocalization takes over, SGP gives rise to greater localization in the Hilbert space than the ideal case up to all accessible evolution times.}
	\label{fig:Z2LGT_comparison} 
\end{figure}

Let us now compare the stabilization of DFL between using the \textit{staggered} protection term with $c_j=[6(-1)^j+5]/11$ in Eq.~\eqref{eq:LPGpro} as has been done in Ref.~\cite{Halimeh2021enhancing}, and employing the SGP term
\begin{align}\label{eq:Z2LGT_SGP}
    V\hat{H}_\mathrm{SGP}=V\sum_jj\big[\hat{W}_j(g^\text{tar}_j)-g^\text{tar}_j\big].
\end{align}
We shall use $g^\text{tar}_j=+1,\,\forall j$, in Eqs.~\eqref{eq:LPGpro} and~\eqref{eq:Z2LGT_SGP} regardless of whether we start in a gauge-invariant state residing in that gauge sector, like $\ket{\psi_0^x}$ in Fig.~\ref{fig:Z2LGT_InitialStates}, or in an initial state that is a superposition over gauge sectors, like $\ket{\psi_0^z}$ in Fig.~\ref{fig:Z2LGT_InitialStates}. We have checked that the chosen value of $g^\text{tar}_j$ does not alter our conclusions. We now prepare our system in $\ket{\psi_0^z}$ of Fig.~\ref{fig:Z2LGT_InitialStates}, which is a domain wall from the perspective of the matter fields, with its left half fully occupied while its right half is empty, but its electric fields are pointed along the $z$-direction such that it becomes a superposition of an extensive number of gauge sectors. We then quench with the faulty theory $\hat{H}_0+\lambda\hat{H}_1+V\hat{H}_\text{pro}$, with $\hat{H}_\text{pro}$ being either Eq.~\eqref{eq:LPGpro} with $c_j=[6(-1)^j+5]/11$ (staggered protection) or the SGP of Eq.~\eqref{eq:Z2LGT_SGP}. The dynamics of the imbalance~\eqref{eq:imbalance} is shown in Fig.~\ref{fig:Z2LGT_comparison}, calculated in ED. In the ideal case of $\lambda=V=0$, we see that the imbalance does not go to zero, settling into a finite-value plateau up to all accessible times. Despite $\ket{\psi_0^z}$ being a translation-invariant state quenched by a translation-invariant nonintegrable model, DFL arises and the system retains memory of its initial state up to all times. Upon accounting for the experimentally unavoidable errors ($\lambda\neq0$), we see in the unprotected case that the imbalance quickly decays to zero, indicating thermalization. Upon adding the staggered protection at the experimentally feasible strength $V=10J$, we see a short trace of enhanced DFL, which then quickly thereafter decays to zero. However, employing SGP at the same protection strength gives rise to enhanced DFL that lasts up to all evolution times accessible in ED.

\begin{figure}[t!]
	\centering
	\includegraphics[width=.48\textwidth]{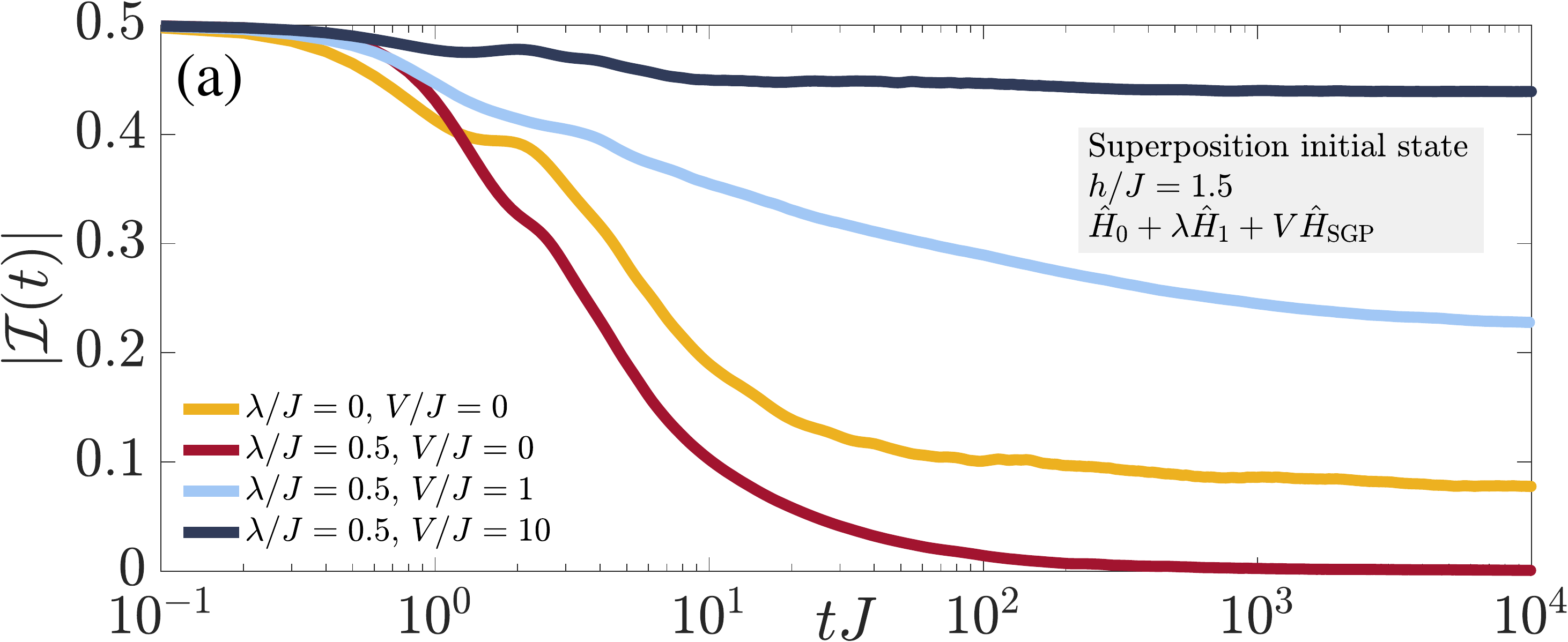}\\
	\vspace{2.1mm}
	\includegraphics[width=.48\textwidth]{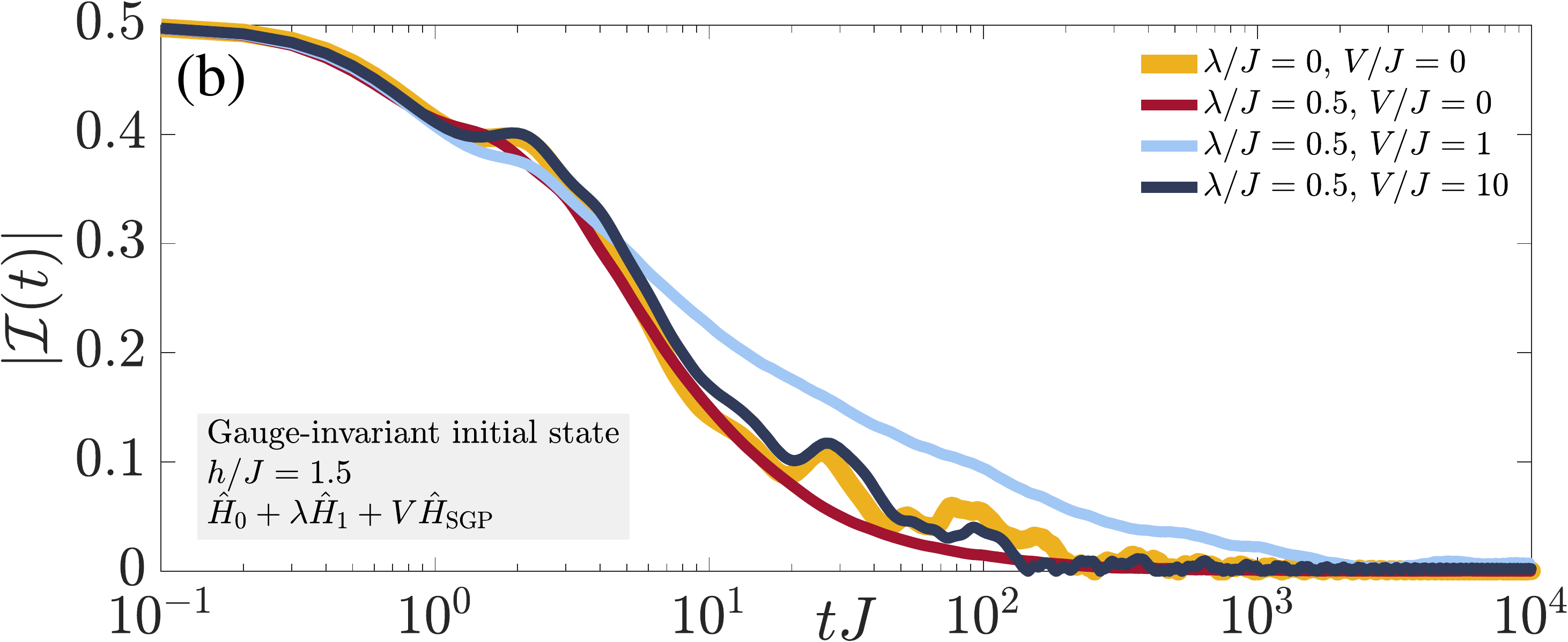}
	\caption{(Color online). Quench dynamics under the faulty theory $\hat{H}_0+\lambda\hat{H}_1+V\hat{H}_\text{SGP}$, where $\hat{H}_0$ is the Hamiltonian~\eqref{eq:Z2LGT} of the $\mathbb{Z}_2$ LGT and $\hat{H}_\text{SGP}$ is the corresponding Stark gauge protection~\eqref{eq:Z2LGT_SGP}. The error term is given in Eq.~\eqref{eq:Z2LGT_H1}. These numerical results are obtained using Krylov-based methods for chains of $L=8$ matter sites and $L=8$ gauge links with periodic boundary conditions. (a) The quench dynamics of the imbalance for the initial state $\ket{\psi_0^z}$ (see Fig.~\ref{fig:Z2LGT_InitialStates}), whose left half is occupied with bosons at each matter site, while its right half is empty, and whose electric fields all point along the $z$-direction, rendering this initial state a superposition of an extensive number of gauge superselection sectors. In the ideal case of $\lambda=V=0$, the dynamics is nonergodic, and disorder-free localization manifests as a finite-value plateau in the imbalance that lasts up to all accessible evolution times. Unprotected errors ($\lambda=0.5J$, $V=0$) quickly destroy DFL. Upon introducing SGP with the local pseudogenerator, Eq.~\eqref{eq:Z2LGT_SGP}, DFL is restored and even enhanced already for a small value of the protection strength $V=J$. For $V=10J$, we observe a stable plateau with a relatively large imbalance. This enhancement in DFL is due to the local pseudogenerator being associated with a local symmetry richer than the $\mathbb{Z}_2$ gauge symmetry of the ideal theory (see text for details). (b) Starting in a gauge-invariant domain-wall state like $\ket{\psi_0^x}$ of Fig.~\ref{fig:Z2LGT_InitialStates} leads to a vanishing imbalance regardless of the values of $\lambda$ and $V$ that we use, and the system thermalizes already over the timescales we simulate.} 
	\label{fig:Z2LGT_Krylov} 
\end{figure}

This enhanced DFL can also be witnessed in the mid-chain entanglement entropy $\mathcal{S}_{L/2}(t)$, shown in Fig.~\ref{fig:Z2LGT_comparison}(b). Unprotected errors lead to further spreading of the wave function in the Hilbert space relative to the ideal case. Upon adding the staggered protection at $V=10J$, localized behavior can be seen at relatively short times, after which the behavior is very close to the case of unprotected errors. However, SGP at the same value of $V=10J$ shows enhanced localization in the Hilbert space up to all accessible evolution times in ED. It is worth remarking here that upon adding the SGP term, we see a faster growth in the mid-chain entanglement entropy at very early times compared to the ideal case in Fig.~\ref{fig:Z2LGT_comparison}(b). This can be explained by looking at the effective Zeno Hamiltonian in the limit of large $V$, which is valid at these early times \cite{Halimeh2021enhancing}:
\begin{align}\label{eq:HQZE}
    \hat{H}_\mathrm{QZE}=V\hat{H}_\mathrm{SGP}+\sum_\mathbf{w}\hat{\mathcal{P}}_\mathbf{w}\big(\hat{H}_0+\lambda\hat{H}_1\big)\hat{\mathcal{P}}_\mathbf{w},
\end{align}
where $\hat{\mathcal{P}}_\mathbf{w}$ are the projectors onto the superselection sectors $\mathbf{w}$ associated with the local symmetry due to the LPG $\hat{W}_j$. The Hamiltonian~\eqref{eq:HQZE} commutes with the LPG, but $\hat{H}_0$ of Eq.~\eqref{eq:Z2LGT} does not. As such, whereas in the ideal case $\hat{H}_0$ can only drive intra-sector dynamics within each sector $\mathbf{g}$, in the enhanced model~\eqref{eq:HQZE} it additionally drives intra- and inter-sector dynamics in the emergent sectors $\mathbf{w}$, which leads to a faster growth of $\mathcal{S}_{L/2}(t)$ at very early times. Nevertheless, when considering the whole period of accessible evolution times in ED, we find that the dynamics exhibits clear DFL behavior that at long times is more localized than under the ideal theory~\eqref{eq:Z2LGT}.

Even though the results of Fig.~\ref{fig:Z2LGT_comparison} are for the $\mathbb{Z}_2$ LGT on a chain of $L=4$ matter sites and $L=4$ gauge links with periodic boundary conditions, our conclusions carry on to larger system sizes. Using Krylov-based time-evolution methods, we repeat the quench with $\hat{H}_0+\lambda\hat{H}_1+V\hat{H}_\text{SGP}$ for a system size of $L=8$ matter sites and $L=8$ gauge links with periodic boundary conditions. The corresponding imbalance dynamics are shown in Fig.~\ref{fig:Z2LGT_Krylov}(a) for the superposition domain-wall initial state $\ket{\psi_0^z}$. Once again, we see that SGP performs quantitatively even better for larger systems, exhibiting a larger value in the resulting DFL plateau compared to Fig.~\ref{fig:Z2LGT_comparison}(a) at $V=10J$. Indeed, even for a small protection-strength value of $V=J$, the imbalance decays very slowly and does not vanish over the timescales we achieve in our numerical simulations.

In order to confirm that the DFL we observe in the presence of SGP is not traditional Stark-MBL, we consider the domain-wall initial state $\ket{\psi_0^x}$ shown in Fig.~\ref{fig:Z2LGT_InitialStates}, but for $L=8$ matter sites and $L=8$ gauge links. This state is gauge-invariant and resides in the gauge superselection sector $g^\text{tar}_j=+1,\,\forall j$. Upon quenching this initial state with $\hat{H}_0+\lambda\hat{H}_1+V\hat{H}_\text{SGP}$, we see that regardless of the considered values of $\lambda$ and $V$, the imbalance decays to zero, indicating that the system thermalizes. This suggests that SGP does not cause localization on its own, but rather that for DFL to occur, the initial state must still be a superposition over an extensive number of gauge sectors. It is interesting to note here that the dynamics under SGP does not quantitatively reproduce the thermalizing dynamics of the ideal case of quenching $\ket{\psi_0^x}$ by Eq.~\eqref{eq:Z2LGT}. This is because the SGP protection at sufficiently large $V$ induces an effective model with an enhanced local symmetry, and therefore the dynamics are expected to be quantitatively different from that of the ideal case, albeit qualitatively similar insomuch that thermalization occurs in both cases.

Finally, we note that we have checked that our conclusions remain the same for different values of $h$ in Eq.~\eqref{eq:Z2LGT} and of the coefficients $\alpha_n$ in Eq.~\eqref{eq:Z2LGT_H1}.

\section{Conclusion and outlook}\label{sec:conc}
In summary, we have introduced the concept of \textit{Stark gauge protection}, which comprises a linear sum of the gauge-symmetry generator or pseudogenerator weighted by coefficients proportional to the matter-site index (Stark potential) associated with the (pseudo)generator. Using exact diagonalization and Krylov-based time-evolution methods, we have shown how it vastly outperforms previous methods in terms of stabilizing and enhancing disorder-free localization, with the latter restored up to all accessible evolution times at moderate values of the protection strength.

Through a Magnus expansion, the details of which can be found in Appendix~\ref{app:ME}, we have shown that by employing Stark gauge protection, the dynamics is effectively propagated by an emergent Hamiltonian in which gauge-breaking terms are suppressed not only by the protection strength, but also by the matter-site index. This suggests that with increasing system size, Stark gauge protection should maintain not just its qualitative but also its quantitative performance. Numerical simulations carried out using Krylov-based time-evolution methods confirm this picture and even show that the performance of Stark gauge protection improves with system size.

We have demonstrated our work on two paradigmatic models, a spin-$S$ $\mathrm{U}(1)$ quantum link model and a $\mathbb{Z}_2$ lattice gauge theory consisting of gauge fields coupled to dynamic matter, both of which are prime models when it comes to modern synthetic quantum matter realizations of gauge theories. Our scheme is readily amenable for implementation in setups of Rydberg atoms with optical tweezers in the case of the $\mathbb{Z}_2$ lattice gauge theory \cite{Halimeh2021enhancing}, and in large-scale Bose--Hubbard quantum simulators for the spin-$1/2$ $\mathrm{U}(1)$ quantum link model \cite{Yang2020,Zhou2021}.

\begin{acknowledgments}
We are grateful to Monika Aidelsburger, Annabelle Bohrdt, Lukas Homeier, Pablo Sala, Christian Schweizer, and Hongzheng Zhao for discussions and work on related projects. This project has received funding from the European Research Council (ERC) under the European Union’s Horizon 2020 research and innovation programm (Grant Agreement no 948141) — ERC Starting Grant SimUcQuam, and by the Deutsche Forschungsgemeinschaft (DFG, German Research Foundation) under Germany's Excellence Strategy -- EXC-2111 -- 390814868. This work is part of and supported by Provincia Autonoma di Trento, the ERC Starting Grant StrEnQTh (project ID 804305), the Google Research Scholar Award ProGauge, and Q@TN — Quantum Science and Technology in Trento. We acknowledge support from the Imperial--TUM flagship partnership. The research is part of the Munich Quantum Valley, which is supported by the Bavarian state government with funds from the Hightech Agenda Bayern Plus.
\end{acknowledgments}

\appendix
\section{Supporting results}\label{app:supp}
\begin{figure}[t!]
	\centering
	\includegraphics[width=.48\textwidth]{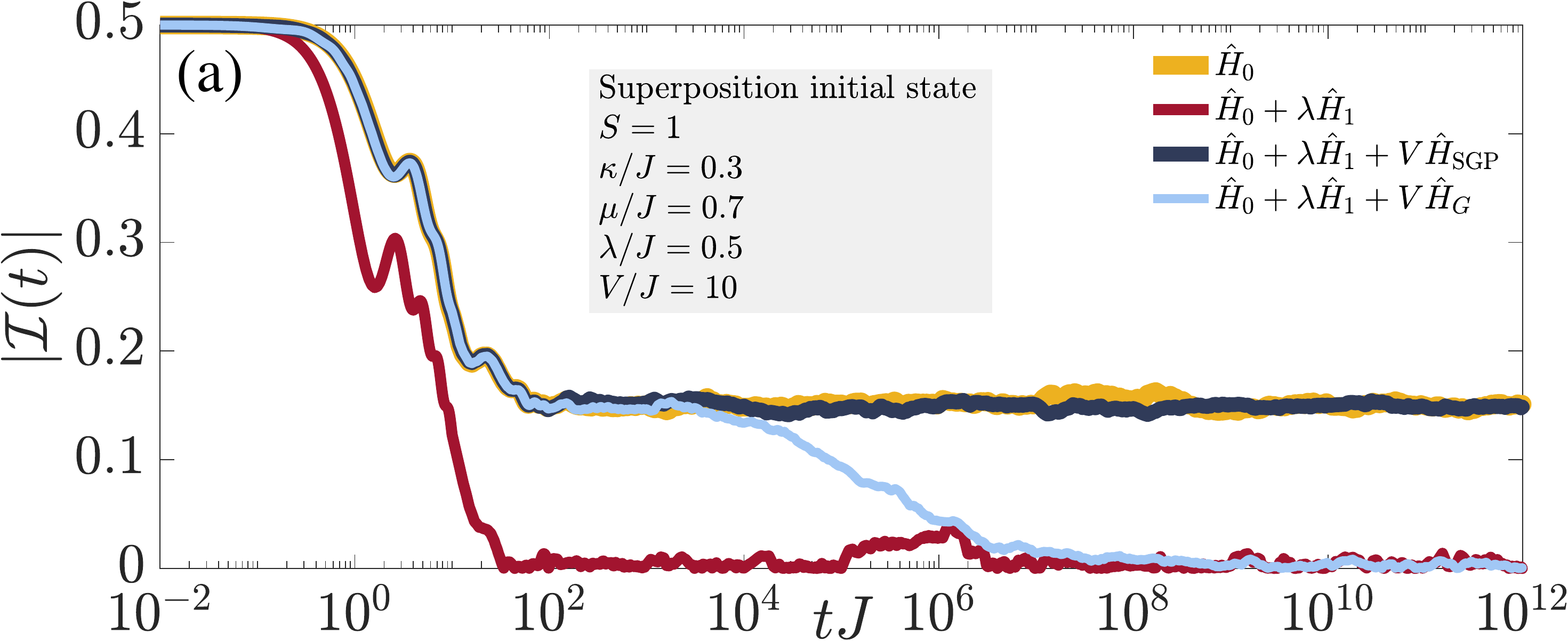}\\
	\vspace{2.1mm}
	\includegraphics[width=.48\textwidth]{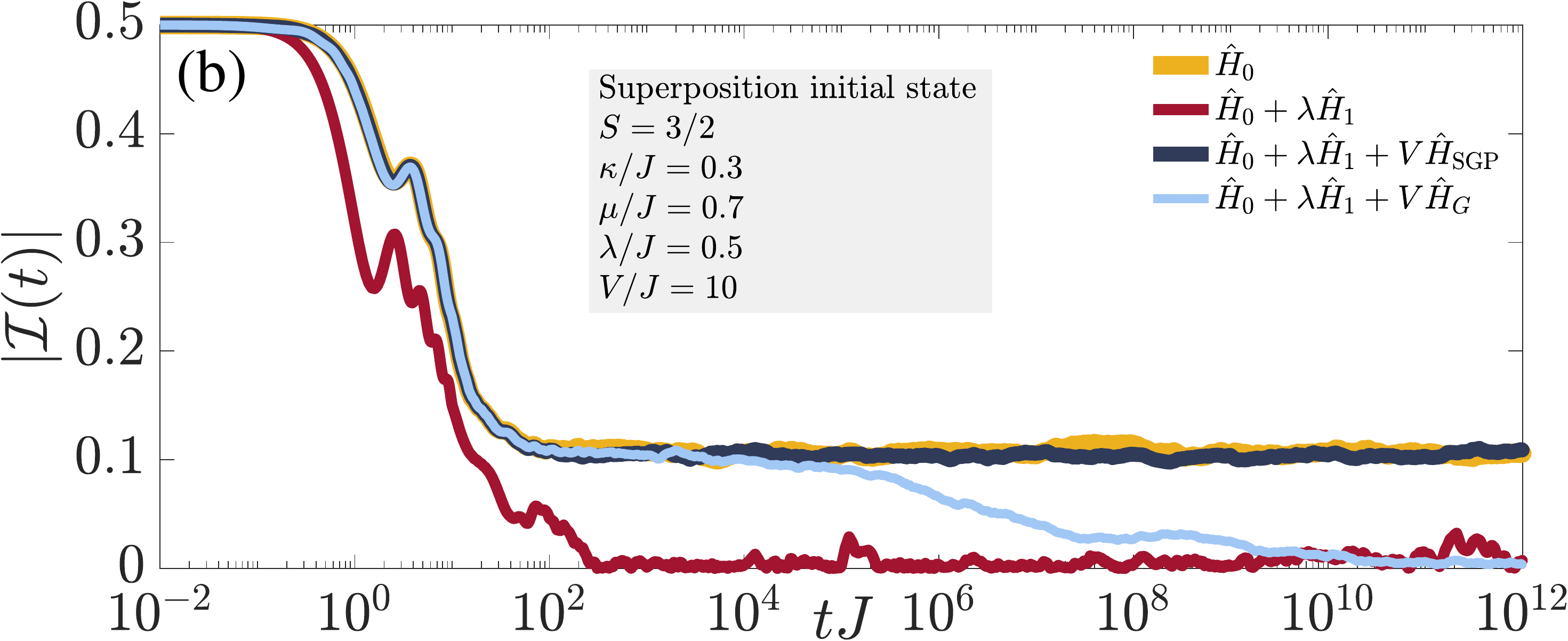}
	\caption{(Color online). Same as Fig.~\ref{fig:U1QLM_comparison}(a), but for a link spin length (a) $S=1$ and (b) $S=3/2$, with an electric-field coupling strength $\kappa/J=0.3$. The performance of SGP is qualitatively the same as in the case of $S=1/2$, stabilizing DFL up to all accessible evolution times.}
	\label{fig:U1QLM_largerS} 
\end{figure}
To demonstrate that the performance of SGP is independent of the link spin length $S$, we have repeated in ED the calculations of Fig.~\ref{fig:U1QLM_comparison}(a) for $S=1$ and $S=3/2$, shown in Fig.~\ref{fig:U1QLM_largerS}(a,b), respectively. The initial state is the same $\ket{\psi_0^x}$ of Fig.~\ref{fig:U1QLM_InitialStates}, which also in the case of $S>1/2$ is a superposition of an extensive number of gauge superselection sectors. We have chosen the strength of the electric-field coupling to be $\kappa/J=0.3$, although we have checked that our conclusions are unchanged for different values of this parameter. 

As can be seen in Fig.~\ref{fig:U1QLM_largerS}, the qualitative behavior of SGP compared with staggered protection is the same for $S>1/2$ as in the case of $S=1/2$, in that whereas the latter protects DFL up to times at most quadratic in the protection strength, SGP stabilizes DFL up to all accessible times in ED.

\begin{widetext}
\section{Magnus expansion}\label{app:ME}
To obtain further insights into the structure of the effective Hamiltonian governing the time evolution at large protection strength $V$, we employ a Magnus expansion \cite{blanes2009magnus}. To this end, 
we divide the \textit{faulty-theory} Hamiltonian into two parts $\hat{H} = V\hat{H}_{\rm pro} + \hat{H}_{\rm bare}$, where $\hat{H}_{\rm bare} = \hat{H}_0 + \lambda\hat{H}_1$, and switch into the interaction picture. The dynamics of an arbitrary operator $\hat{O}$ is \begin{equation}
    \hat{O}(t) = e^{iV\hat{H}_{\rm pro}t}\hat{O}(0)e^{-iV\hat{H}_{\rm pro}t}.
\end{equation}
Due to the integer spectrum of $\hat{H}_{\rm pro}$, $\hat{O}(t)$ is a time-periodic operator $\hat{O}(t) = \hat{O}(t + mT)$, $m\in\mathbb{Z}$, with the period $T = 2\pi/V$. Specifically, the Fourier series of $\hat{H}_{\rm bare}$ is given by
\begin{equation}\label{eq:Fourier}
    \hat{H}_{\rm bare}(t) = \sum_{l = -\infty}^{+\infty} \hat{H}_{{\rm bare},l}e^{ilVt}.
\end{equation}
In the systems considered in this article, 
\begin{equation}
\hat{H}_{{\rm bare},0}\equiv \hat{H}_0, 
\end{equation}
which corresponds to the gauge-invariant ideal Hamiltonian. The terms with $l\neq 0$ contribute to the gauge-breaking processes.

The time-evolution operator for the state is 
\begin{equation} \label{eq:interaction}
    \hat{U}(t) = \mathcal{T}\Big\{e^{-i\int_0^t d\tau\hat{H}_{\rm bare}(\tau)}\Big\},
\end{equation}
with $\mathcal{T}$ the time-ordering operator. The procedure of the Magnus expansion \cite{blanes2009magnus,bukov2015universal} is to re-express Eq.~(\ref{eq:interaction}) as a real exponential of an operator, $\hat{U}(t) = \exp\hat{\Omega}(t)$, where the skew-Hermitian matrix $\hat{\Omega}(t)$ is constructed as a series expansion $\hat{\Omega}(t) = \sum_{n=0}^\infty\hat{\Omega}_n(t)$ in powers $n$ of $\hat{H}_{\rm bare}$. The first three terms of the expansion are
\begin{subequations}
\begin{align}\label{eq:OmegaExpression}
    \hat{\Omega}_0(t) &= -i\int_0^t dt_1\,\hat{H}_{\rm bare}(t_1), \\
    \hat{\Omega}_1(t) &= -\frac{1}{2}\int_0^t dt_1 \int_0^{t_1} dt_2\,\big[\hat{H}_{\rm bare}(t_1),\hat{H}_{\rm bare}(t_2)\big],   \\
    \hat{\Omega}_2(t) &= \frac{i}{6}\int_0^t dt_1 \int_0^{t_1} dt_2 \int_0^{t_2} dt_3\,\Big(\big[\hat{H}_{\rm bare}(t_1),\big[\hat{H}_{\rm bare}(t_2),\hat{H}_{\rm bare}(t_3)\big]\big] + \big[\hat{H}_{\rm bare}(t_3),\big[\hat{H}_{\rm bare}(t_2),\hat{H}_{\rm bare}(t_1)\big]\big] \Big).
\end{align}
\end{subequations}
The absolute convergence of the above expansion requires \cite{blanes2009magnus}
\begin{equation}
    \int_0^T dt\, \big\lvert\big\lvert\hat{H}_{\rm bare}(t)\big\rvert\big\rvert < 0.20925,
\end{equation}
which can be always satisfied for finite systems with a sufficiently large $V\sim JL$.

Equation~(\ref{eq:OmegaExpression}) can be dramatically simplified at the stroboscopic time $t = mT$. For convenience, define $\hat{U}(mT) = \exp\big\{-imT\hat{H}_{\rm eff}\big\}$ and $\hat{H}_{\rm eff} = \sum_{n=0}^\infty \hat{H}_{\rm eff}^{(n)}$. The relation between the $n^\text{th}$-order effective Floquet Hamiltonian $\hat{H}_{\rm eff}^{(n)}$ and $\hat{\Omega}_n$ is $\hat{H}_{\rm eff}^{(n)} = i\hat{\Omega}_n(T)/T$. With the help of the Fourier series Eq.~(\ref{eq:Fourier}), the effective Floquet Hamiltonian can be expressed as \cite{blanes2009magnus,bukov2015universal}
\begin{subequations}
\begin{align}
    \hat{H}_{\rm eff}^{(0)} &= \hat{H}_{{\rm bare},0}=\hat{H}_0,\\\label{eq:FirstOrderEff}
    \hat{H}_{\rm eff}^{(1)} &= \sum_{l=1}^\infty \frac{1}{lV} \Big(\big[\hat{H}_{{\rm bare},l},\hat{H}_{{\rm bare},-l}\big]  - \big[\hat{H}_{{\rm bare},l},\hat{H}_{{\rm bare},0}\big] + \big[\hat{H}_{{\rm bare},-l},\hat{H}_{{\rm bare},0}\big] \Big).
\end{align}
\end{subequations}
When $V$ is sufficiently large, the dominant Hamiltonian is $\hat{H}_{\rm eff}^{0}$, and $\hat{H}_{\rm eff}^{(1)}$ is negligible, which means that the ideal theory is reliably protected. This coincides with the QZE prediction \cite{Halimeh2021stabilizingDFL,Halimeh2021enhancing}. The first-order effective Hamiltonian~\eqref{eq:FirstOrderEff} contains the gauge-breaking processes $\big[\hat{H}_{{\rm bare},l>0},\hat{H}_{{\rm bare},0}\big]$, which lead to the breakdown of DFL for the case of staggered protection, but, as we will show, not for the case of SGP. Now we consider the $U(1)$ QLM. For notational brevity, we define $\Gamma = J/\big[2\sqrt{S(S+1)}\big]$ (with the lattice spacing set to $a=1$). The explicit form of the effective Floquet Hamiltonian for the staggered protection is 
\begin{align}\nonumber
    \hat{H}_{\rm eff}^{(1)} = & \sum_j \frac{\lambda}{2V}\bigg\{\frac{\lambda}{2}\bigg[ 2\hat{\sigma}_{j-1}^-\hat{\sigma}_{j}^z\hat{\sigma}_{j+1}^+ +  2\hat{\sigma}_{j-1}^+\hat{\sigma}_j^z\hat{\sigma}_{j+1}^- + \frac{\hat{s}^z_{j,j+1}}{S(S+1)} + \hat{\sigma}_j^z + \hat{\sigma}_{j+1}^z \bigg]\\\nonumber
    & - \bigg[ \frac{J\hat{\sigma}_j^-\hat{s}^z_{j,j+1}\hat{\sigma}_{j+1}^-}{2S(S+1)} - 2\mu\hat{\sigma}_j^-\hat{\sigma}_{j+1}^-+\frac{\Gamma}{2}\hat{\sigma}_j^z\hat{s}_{j,j+1}^-
    +\frac{\Gamma}{2}\hat{s}_{j,j+1}^-\hat{\sigma}_{j+1}^z + \Gamma
    \hat{\sigma}_j^-\hat{\sigma}_{j+1}^z\hat{s}^-_{j+1,j+2}\hat{\sigma}_{j+2}^+ \\\label{eq:ME_staggered}
    &+ \Gamma\hat{\sigma}_{j-1}^+\hat{s}^-_{j-1,j}\hat{\sigma}_{j}^z\hat{\sigma}_{j+1}^- - \frac{\kappa^2 }{4\sqrt{S(S+1)}}\big(\hat{s}_{j,j+1}^z\hat{s}_{j,j+1}^+ + \hat{s}_{j,j+1}^+\hat{s}_{j,j+1}^z\big) + \rm{H.c.} \bigg]\bigg\}.
\end{align}

For the SGP term, $\hat{H}_{\rm eff}^{(1)}$ also contains gauge-breaking processes. In principle, these terms could also lead to the breakdown of DFL. However, the numerical results show the stabilization of DFL for all the accessible evolution times in ED and Krylov-based methods.  As we discuss in the main text, this arises from the effective inhomogeneous Hamiltonian, which differs from the well-known Hilbert space fragmentation and Stark-MBL. The explicit form of $\hat{H}_{\rm eff}^{(1)}$ in the case of SGP is
\begin{align}
   \hat{H}_{\rm eff}^{(1)} = &\frac{\lambda^2}{2(L+1)V}\bigg[  \frac{\hat{s}^z_{L,L+1}}{S(S+1)} + \hat{\sigma}_L^z + \hat{\sigma}_{L+1}^z \bigg] +\sum_{j\ne L} \frac{\lambda^2}{2(2j+1)V}\bigg[  \frac{\hat{s}^z_{j,j+1}}{S(S+1)} + \hat{\sigma}_j^z + \hat{\sigma}_{j+1}^z \bigg] \nonumber \\
    &- \frac{\lambda}{(L+1)V}\bigg[  \frac{J\hat{\sigma}_L^-\hat{s}^z_{L,L+1}\hat{\sigma}_{L+1}^-}{2S(S+1)} - 2\mu\hat{\sigma}_L^-\hat{\sigma}_{L+1}^- +\frac{\Gamma}{2}\hat{\sigma}_L^z\hat{s}_{L,L+1}^-+\frac{\Gamma}{2}\hat{s}_{L,L+1}^-\hat{\sigma}_{L+1}^z\nonumber \\ 
    & + \Gamma\hat{\sigma}_L^-\hat{\sigma}_{L+1}^z\hat{s}^-_{L+1,L+2}\hat{\sigma}_{L+2}^+ + \Gamma\hat{\sigma}_{L-1}^+\hat{s}^-_{L-1,L}\hat{\sigma}_{L}^z\hat{\sigma}_{L+1}^-  - \frac{\kappa^2 a}{4\sqrt{S(S+1)}}\hat{s}_{L,L+1}^z\hat{s}_{L,L+1}^+ \nonumber \\
    &- \frac{\kappa^2 a}{4\sqrt{S(S+1)}}\hat{s}_{L,L+1}^+\hat{s}_{L,L+1}^z + {\rm{H.c.}} \bigg] 
    - \sum_{j\ne L} \frac{\lambda}{(2j+1)V}\bigg[ \frac{J\hat{\sigma}_j^-\hat{s}^z_{j,j+1}\hat{\sigma}_{j+1}^-}{2S(S+1)}- 2\mu\hat{\sigma}_j^-\hat{\sigma}_{j+1}^- \nonumber \\
    & +\frac{\Gamma}{2}\hat{\sigma}_j^z\hat{s}_{j,j+1}^- +\frac{\Gamma}{2}\hat{s}_{j,j+1}^-\hat{\sigma}_{j+1}^z +\Gamma\hat{\sigma}_j^-\hat{\sigma}_{j+1}^z\hat{s}^-_{j+1,j+2}\hat{\sigma}_{j+2}^+  + \Gamma\hat{\sigma}_{j-1}^+\hat{s}^-_{j-1,j}\hat{\sigma}_{j}^z\hat{\sigma}_{j+1}^-  \nonumber \\\label{eq:ME_SGP}
    &- \frac{\kappa^2 }{4\sqrt{S(S+1)}}\hat{s}_{j,j+1}^z\hat{s}_{j,j+1}^+ 
    - \frac{\kappa^2 }{4\sqrt{S(S+1)}}\hat{s}_{j,j+1}^+\hat{s}_{j,j+1}^z 
    +\rm{H.c.} \bigg].
\end{align}
Note how in both Eqs.~\eqref{eq:ME_staggered} and~\eqref{eq:ME_SGP} the terms $\propto\kappa^2$ automatically cancel each other out in the case of $S=1/2$. This makes sense because the term $\propto\big(\hat{s}^z_{j,j+1}\big)^2$ in Eq.~\eqref{eq:H0} is an inconsequential constant energy term when $S=1/2$ that can be omitted. 

Let us now numerically investigate the resulting effective Hamiltonian
\begin{align}\label{eq:effective}
    \hat{H}_\mathrm{eff}=\hat{H}_0+\hat{H}_\mathrm{eff}^{(1)},
\end{align}
with $\hat{H}_\mathrm{eff}^{(1)}$ given in Eq.~\eqref{eq:ME_SGP}, for the spin-$1/2$ $\mathrm{U}(1)$ QLM. We first consider the superposition initial state $\ket{\psi_0^x}$ shown in Fig.~\ref{fig:U1QLM_InitialStates}. As we have already established, its quench dynamics under $\hat{H}_0$ will lead to DFL, with the imbalance settling into a nonzero plateau up to all accessible times. Even though gauge-breaking errors $\lambda\hat{H}_1$ with $\lambda\neq0$ will destroy DFL, adding SGP and quenching by $\hat{H}_0+\lambda\hat{H}_1+V\hat{H}_\mathrm{SGP}$ at moderate or large $V$ will stabilize DFL up to all times; see Fig.~\ref{fig:U1QLM_comparison}(a). We show this in Fig.~\ref{fig:ME}(a) for $\lambda=0.5J$ and $V=100J$. We now compare this to the imbalance under the effective Hamiltonian $\hat{H}_\text{eff}$~\eqref{eq:effective}. As shown in Fig.~\ref{fig:ME}(a), DFL is restored up to all times, with quantitative agreement with the corresponding case of $\hat{H}_0+\lambda\hat{H}_1+V\hat{H}_\mathrm{SGP}$ up to times at least $\propto V^2/(J^2\lambda)$ (see below). Interestingly, when we trivially remove the site-dependence in the prefactors of $\hat{H}_\text{eff}$, we get a spatially homogeneous Hamiltonian $\hat{H}_\text{eff}'$. The imbalance under $\hat{H}_\text{eff}'$ looks qualitatively very similar to the case of a quench by $\hat{H}_0+\lambda\hat{H}_1+V\hat{H}_G$, i.e., under staggered protection. The imbalance is stabilized up to times at most quadratic in $V$, after which it decays to zero due to thermalization. This strongly suggests that the site-dependence of the prefactors in the effective Hamiltonian~\eqref{eq:effective} is the key ingredient of the permanent stabilization of DFL in the case of SGP.

For completeness, we also check the dynamics of the imbalance under the effective Hamiltonian~\eqref{eq:effective} after starting in the gauge-invariant initial state $\ket{\psi_0^z}$ of Fig.~\ref{fig:U1QLM_InitialStates}. The corresponding result is shown in Fig.~\ref{fig:ME}(b), where we again find that the imbalance decays to zero, and shows quantitative agreement with the case of SGP up to times at least $\propto V^2/(J^2\lambda)$, as we have checked numerically.

\begin{figure*}[t!]
	\centering
	\includegraphics[width=.48\textwidth]{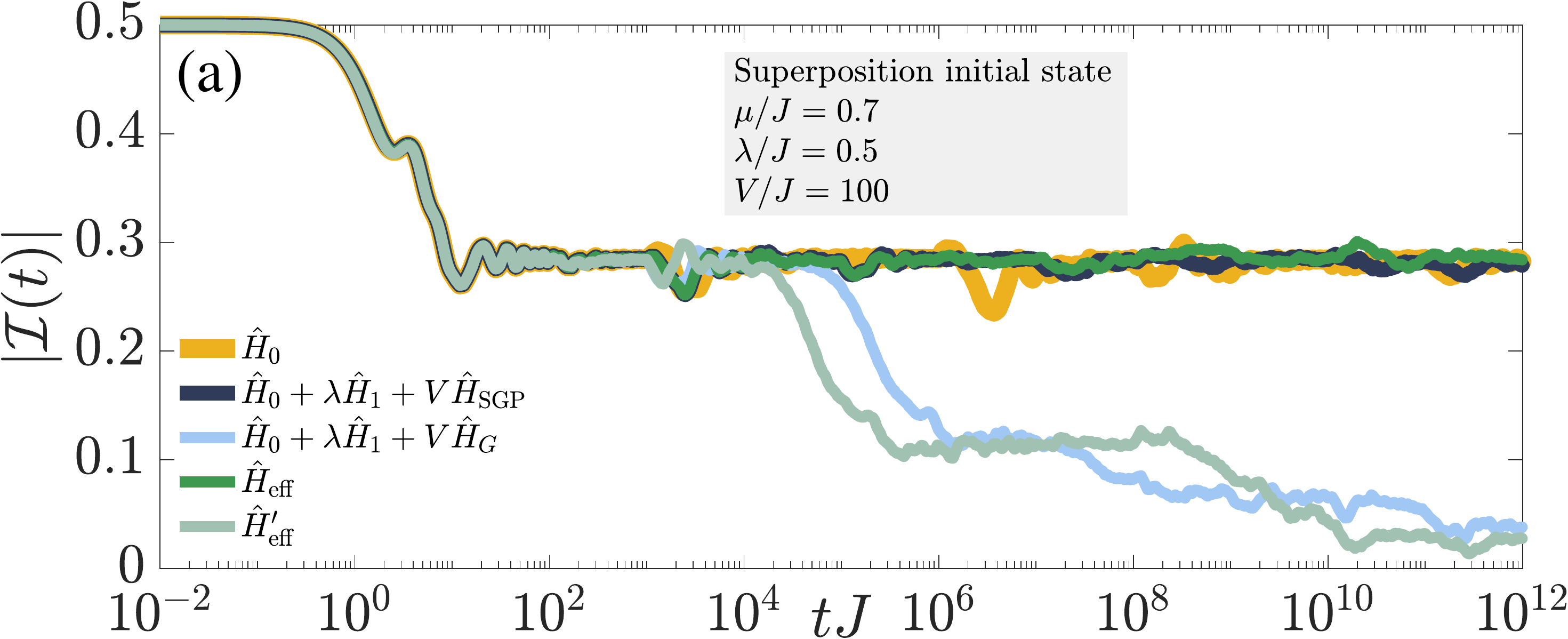}\quad
	\vspace{2.1mm}
	\includegraphics[width=.48\textwidth]{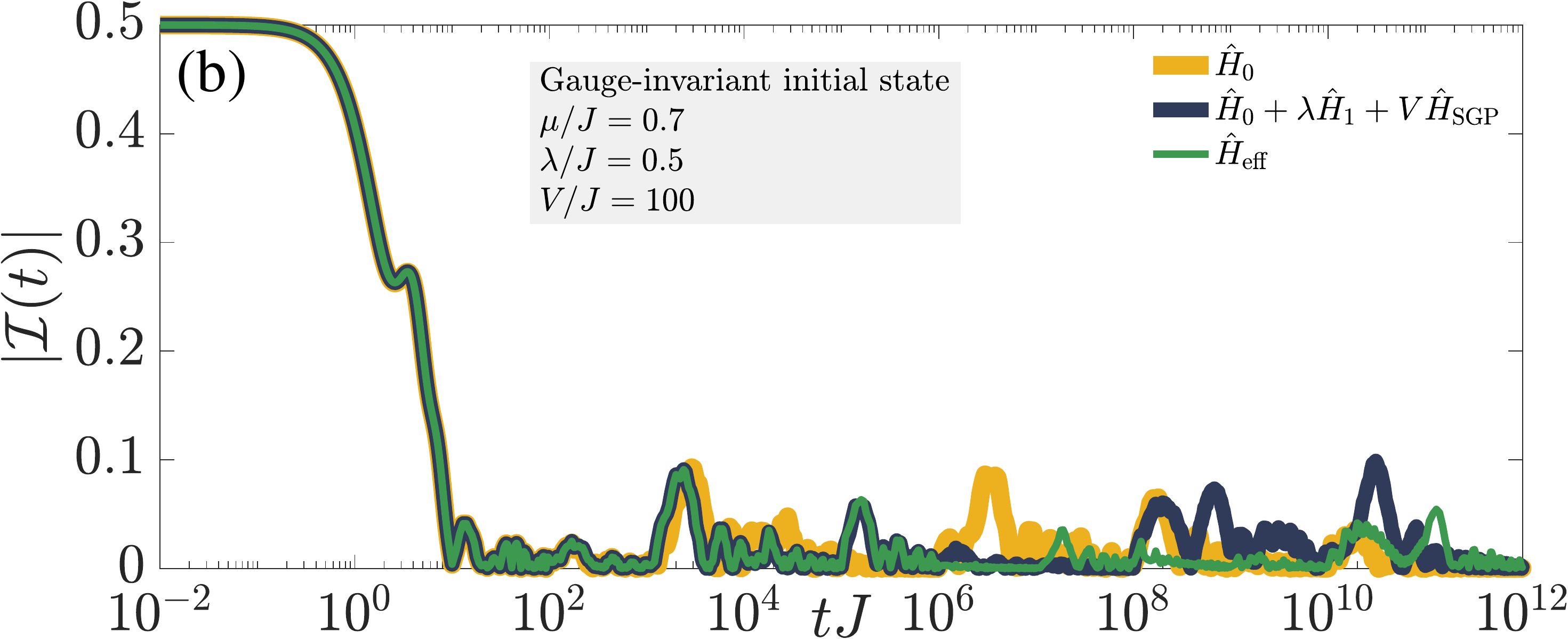}
	\caption{(Color online). Numerically investigating the effective Hamiltonian (up to first-order in Floquet theory) given in Eq.~\eqref{eq:effective}. (a) In the case of the superposition initial state $\ket{\psi_0^x}$ (see Fig.~\ref{fig:U1QLM_InitialStates}), it leads to a finite imbalance plateau up to all accessible evolution times, showing very good quantitative agreement with the faulty theory $\hat{H}_0+\lambda\hat{H}_1+V\hat{H}_\mathrm{SGP}$ up to at least $t\propto V^2/(J^2\lambda)$, as we have numerically checked. When trivially removing the site-dependence of the prefactors in Eq.~\eqref{eq:ME_SGP}, the imbalance looks qualitatively similar to the case of staggered protection (i.e., a quench under $\hat{H}_0+\lambda\hat{H}_1+V\hat{H}_G$), with the DFL restored up to times quadratic in $V$, after which the imbalance decays and the system thermalizes. (b) In the case of the gauge-invariant initial state $\ket{\psi_0^z}$, we also find that the effective Hamiltonian~\eqref{eq:effective} gives the same qualitative result as the ideal theory and the faulty theory under SGP, where the imbalance decays to zero. There is also very good quantitative agreement between the case with SGP and that with the effective Hamiltonian up to times at least quadratic in $V$, as we have checked numerically.}
	\label{fig:ME} 
\end{figure*}

Let us now determine the timescale up to which $\hat{H}_{\rm eff}$~\eqref{eq:effective} is expected to quantitatively agree with the faulty theory $\hat{H}_0+\lambda\hat{H}_1+V\hat{H}_\mathrm{SGP}$. In order to do this, let us look at the second-order effective Floquet Hamiltonian, which reads
\begin{align}\nonumber
    \hat{H}_{\rm eff}^{(2)} =& -\sum_{l_1,l_2,l_3 = -\infty}^{+\infty}\frac{1}{6T}\int_0^T dt_1 \int_0^{t_1} dt_2 \int_0^{t_2} dt_3\big[\hat{H}_{{\rm bare},l_1},\big[\hat{H}_{{\rm bare},l_2},\hat{H}_{{\rm bare},l_3}\big]\big]\\\label{eq:SecondH}
    &\times\Big( e^{il_1Vt_1 + il_2Vt_2 + il_3Vt_3} + e^{il_1Vt_3 + il_2Vt_2 + il_3Vt_1}\Big).
\end{align}
When $V \ll J\sim\mu \ll \lambda$, the largest contribution in the second-order effective Floquet Hamiltonian Eq.~\eqref{eq:SecondH} is given by
\begin{align}
    &-\frac{1}{6T}\int_0^T dt_1 \int_0^{t_1} dt_2 \int_0^{t_2} dt_3\big[\hat{H}_{{\rm bare},0},\big[\hat{H}_{{\rm bare},0},\hat{H}_{{\rm bare},l}\big]\big]\Big( e^{ilVt_1} + e^{ilVt_3} - 2 e^{ilVt_2} \Big) \nonumber \\
    &= -\frac{1}{V^2l^2}\big[\hat{H}_{{\rm bare},0},\big[\hat{H}_{{\rm bare},0},\hat{H}_{{\rm bare},l}\big]\big]\sim \frac{J^2L\lambda}{V^2},
\end{align}
which yields the ``worst-case'' valid timescale $t\sim V^2/(J^2\lambda L)$ for the Hamiltonian $\hat{H}_{\rm eff}=\hat{H}_0 + \hat{H}_{\rm eff}^{(1)}$. We have checked numerically that this is indeed the case.
\end{widetext}

\bibliography{SGP_biblio}
\end{document}